\documentclass[preprint]{elsarticle}
\usepackage[utf8]{inputenc}
\usepackage{amsmath,rotating}
\usepackage{dashrule}
\usepackage[margin=3cm]{geometry}
\usepackage{xcolor}
\usepackage{hyperref}
\usepackage[capitalize]{cleveref}
\usepackage{textcomp}
\usepackage{setspace}
\usepackage{etoolbox}
\patchcmd{\pprintMaketitle}
{\hrule\vskip12pt}
{\hrule\vskip12pt\ifvoid\extrainfobox\else\unvbox\extrainfobox\par\vskip12pt\fi}
{}{}

\newenvironment{extrainfo}
{\global\setbox\extrainfobox=\vbox\bgroup\parindent=0pt }
{\egroup}
\newsavebox\extrainfobox

\makeatletter
\def\ps@pprintTitle{%
	\let\@oddhead\@empty
	\let\@evenhead\@empty
	\let\@oddfoot\@empty
	\let\@evenfoot\@oddfoot
}
\makeatother

\newcommand{\pdts}[1]{\partial_t #1}
\newcommand{\pdt}[1]{\partial_t\left(#1\right)}
\newcommand{\pdxs}[1]{\partial_x #1}
\newcommand{\pdx}[1]{\partial_x\left(#1\right)}

\newcommand{\vavg}[1]{\overline{#1}}
\newcommand{\phavg}[1]{\langle #1 \rangle}
\newcommand{\favg}[1]{\tilde{#1}}
\newcommand{\wfavg}[1]{\widetilde{#1}}

\newcommand{\fui}{\favg{u}_i}
\newcommand{\fuj}{\favg{u}_j}

\newcommand{\fu}{\favg{u}}
\newcommand{\phrho}{\phavg{\rho}}

\newcommand{\php}{\phavg{p}}
\newcommand{\fe}{\favg{e}}

\newcommand{\Rep}{\mathrm{Re_p}}
\newcommand{\Ma}{\mathrm{Ma}}
\newcommand{\Map}{\mathrm{Ma_p}}
\newcommand{\is}{\mathrm{IS}}
\newcommand{\Vp}{V_\mathrm{p}}
\newcommand{\Dp}{D_\mathrm{p}}
\newcommand{\Rp}{R_\mathrm{p}}
\newcommand{\supun}{^{\mathrm{un}}}
\newcommand{\alphap}{\alpha_\mathrm{p}}
\newcommand{\alphasep}{\alpha_\mathrm{sep}}
\newcommand{\taup}{\tau_\mathrm{p}}
\newcommand{\Fqs}{F_\mathrm{qs}}
\newcommand{\Fiu}{F_\mathrm{iu}}
\newcommand{\Fvu}{F_\mathrm{vu}}

\newcommand{\Fun}{F_\mathrm{un}}

\usepackage{ulem}

\begin{document}
\begin{frontmatter}

\title{Performance of drag force models for shock-accelerated flow in dense particle suspensions}

\author[a]{Andreas Nygård Osnes\corref{cor1}}
\cortext[cor1]{Corresponding author}
\ead{Andreas-Nygard.Osnes@ffi.no}

\author[a]{Magnus Vartdal}
\ead{Magnus.Vartdal@ffi.no}
\address[a]{Norwegian Defence Research Establishment, PO Box 25 Kjeller, 2027 Kjeller, Norway}

\journal{}

\begin{extrainfo}
	This is the author's accepted manuscript for the original manuscript to be published in International Journal of Multiphase Flow.\\
	This manuscript version is made available under the CC-BY-NC-ND 4.0 license\\ \url{https://creativecommons.org/licenses/by-nc-nd/4.0/}
\end{extrainfo}

\begin{abstract}
Models for prediction of drag forces within a particle cloud following shock-acceleration are evaluated with the aid of results from particle-resolved simulations in order to quantify how much the disturbances introduced by the proximity of nearby particles affect the drag forces. The drag models evaluated here consist of quasi-steady forces, undisturbed flow forces, inviscid unsteady forces, and viscous unsteady forces. Two dense particle curtain correction schemes to these forces, based on volume fraction and input velocity, are also evaluated. The models are tested in two ways; first they are evaluated based on volume-averaged flow fields from particle-resolved simulations; secondly, they are applied in Eulerian-Lagrangian simulations, and the results are compared to the particle-resolved simulations.  

The results show that both correction schemes significantly improve the particle force predictions, but the average total impulse on the particles is still underpredicted by both correction schemes in both tests. With the volume averaged flow fields as input, the volume fraction correction gives the best results. However, in the Eulerian-Lagrangian simulations it is demonstrated that the velocity fluctuation model, associated with the velocity correction scheme, is crucial for obtaining accurate predictions of the mean flow fields.   
	
\end{abstract}
\end{frontmatter}
\section{Introduction}	

Simulations of particle motion require accurate models for the interaction forces between particles and the surrounding fluid flow. While particle-resolved simulations give accurate predictions of drag and particle movement, they are in most cases too computationally demanding to be applicable to full-scale systems of interest. Therefore, it is necessary to use less computationally expensive methods, such as Eulerian-Eulerian (EE) simulations or Eulerian-Lagrangian (EL) simulations. The simplifications made by these models introduce new modeling challenges, since the methods must be supplied with models for the physical processes that occur at scales smaller than the computational grid, such as particle drag.  For isolated particles, there exists a number of drag correlations that are very accurate, and can model the forces reliably over a large range of Mach numbers, Reynolds numbers etc. However, in the presence of other particles, it is questionable how well these models are able to capture the particle drag, since they do not account for the fluid mediated particle-particle forces. Additionally, the classical drag models require values for undisturbed fluid properties at the particle location, i.e. the fluid properties that would be observed there in absence of the particle. These properties are generally not available in problems with dense particle suspensions. Therefore, it is nontrivial to compute particle forces in such problems, and while many studies have performed simulations of this kind, e.g. \citep{utkin2017,osnes2018,sugiyama2019,gai2020}, it is unclear whether the physical processes at particle scale are well represented, and therefore whether or not the results are reliable. This motivates a detailed assessment of the performance of classical drag laws in problems featuring dense particle suspensions. The purpose of this work is to assess how well classical drag laws can represent the particle drag in the setting of shock-accelerated flow through a layer of stationary particles at 10\% volume fraction. 

To characterize how well the drag laws can predict the drag forces, we use data from particle-resolved Large Eddy Simulations of a Mach 2.6 shock wave propagating through a stationary particle layer with an initial particle Reynolds number of 2000. The data from these simulations were analyzed in \citet{osnes2019c}. In that work, it was demonstrated that direct application of standard drag-laws underpredicted the late-time particle forces by up to 50\%, and performed worse with increasing particle Reynolds number. Here, we take two different approaches to identify the applicability of the drag laws. First, we use the flow data and particle forces from the particle-resolved simulations directly, and compare how well the particle forces are predicted by classical models based on the volume-averaged flow properties. Secondly, we perform EL simulations of the same problem, and again compare how well the particle forces are represented. In this second approach, we also compare the volume-averaged flow properties to those observed in the particle-resolved simulations. Since there is a strong two-way coupling between the particles and the fluid flow, the particle force model has a strong influence on the mean fluid flow. 

Particle-resolved simulations of flow through dense particle suspensions have proven to be a valuable approach for understanding the physical processes occurring in the interior of particle suspensions. Such simulations give access to finely resolved temporal and spatial data in regions that are very challenging to probe in experimental studies. Particle-resolved simulations have therefore become popular in recent years. For shock-accelerated flows, three-dimensional simulations were used in \cite{mehta2016,mehta2018,mehta2019,mehta2020,theofanous2018,vartdal2018,osnes2019,osnes2019c}. Several two-dimensional particle-resolved simulations have also been conducted, e.g. \cite{regele2014,hosseinzadeh2018}.

The particle-resolved studies have shown that the local particle configuration has a large influence on the particle drag forces \citep{akiki2016,mehta2018,mehta2019}. This is the case both for the inviscid shock-related forces and for the later quasi-steady drag forces. The forces imposed by the shock wave can be amplified or weakened due to shock wave diffraction. Later, after development of particle wakes, there are strong velocity gradients oriented in all directions within the suspension. Therefore, particles can be located in regions of both very high and very low flow speeds, with correspondingly strong or weak drag forces. An additional complicating factor is the possibility of local transonic flow regions, where larger voids in the particle suspension can act as expanding nozzles that accelerate the flow from subsonic to supersonic in a limited region. Shocklet formation around particles in such regions is also possible.

Forces during the initial, primarily inviscid, shock-accelerated flow in dense particle suspensions have been analyzed in \citet{mehta2018,mehta2019}. Peak drag decays with downstream distance, as would be expected due to shock wave attenuation, but the magnitude was also found to depend drastically on statistical properties of the particle configuration; face-centered cubic or simple cubic packings had larger peak drag coefficients than a random distribution on average. Similar configuration effects are also likely to be the case for the quasi-steady drag coefficients, although to the authors knowledge this has not yet been investigated for high-speed flows.

High-speed flow through particle clouds has also been shown to be highly unsteady, and flow fluctuations can be significant in many configurations. These fluctuations consist of both pseudo-turbulent fluctuations and classical turbulent fluctuations. While there is not a clear-cut distinction between these, pseudo-turbulent fluctuations are related to the disturbance flow around a particle, which does not need to be turbulent, but still acts as a Reynolds-stress term under statistical averaging. The kinetic energy contained in such fluctuations can be significant \cite{regele2014,vartdal2018,osnes2019,mehta2020}, and this is important for modeling. For example, the pressure in EL simulations are often computed by subtracting the mean kinetic energy from the total energy and subsequently employing the equation of state, while the correct approach would be to subtract the total kinetic energy. Flow fluctuations also affect the drag-forces, and recent works have characterized the distribution of peak forces, temporal drag-fluctuations, and the temporally averaged forces \citep{mehta2019,osnes2019c}.   

There are several properties of the particle cloud that have effects on the drag forces. These properties include, but are not limited to, the particle volume fraction, the volume fraction gradients, and the local particle configuration. In total, this complicated dependence represents a formidable modeling challenge. However, fluid-mediated particle-particle interaction models are currently actively researched and are able to capture some of these phenomena. Notable studies in this direction are \citet{akiki2017,akiki2017b,sen2018,moore2019,balachandar2020}. 

In this work, we aim to characterize models that are readily implemented in common EL codes in use today, such as the model by \citet{parmar2010}. We also characterize a physics-based correction to the input to the drag models, which was recently proposed by \cite{osnes2019,osnes2019c}. Fluid-mediated particle-particle interaction models require significantly more effort to implement, often by relying on available particle resolved DNS data, and have yet to be extended to the higher flow speeds of interest in this work. Thus they fall outside the scope of the current work. The drag models considered here are also of interest for EE simulation strategies, where detailed particle  configuration data is not available. A recent discussion of the properties of such simulations is found in \citep{fox2020}.

Improved simulation capabilities for high-speed, dense, multi-phase flows are advantageous for a wide range of problems. A few examples are volcanic eruptions \citep{zwick2019}, meteoroid breakup \cite{mcmullan2019}, needle-free drug delivery systems \citep{truong2006}, solid and liquid fuel engines \citep{cai2003,shimada2006,bravo2015}, explosion mitigation or explosive dispersal \cite{zhang2001,milne2010,gottiparthi2014}, and noise attenuation on rocket launch pads \citep{ignatius2008}.

This article is structured as follows. In \cref{sec:prforces}, the particle-resolved simulation data is used as input to the particle force models and the results are compared to the forces obtained in the particle-resolved simulations. \Cref{sec:prsims} briefly describes the particle-resolved simulations, \cref{sec:forcemodels} introduces the different force models. In \cref{sec:singlespheresimulations}, particle-resolved simulations and the force models are applied to two single-sphere problems, while \cref{sec:prforcecomparison} compares the forces of the shock-wave particle-cloud simulations with the force predictions. Next, the force models are evaluated as part of EL simulations in \cref{sec:ELforces}. The EL simulation method is presented in \cref{sec:ELmethod}, while the simulation results and comparison to the particle-resolved simulations are contained in \cref{sec:ELsimulations}. Finally, \cref{sec:conclusions} contains concluding remarks and a discussion of possible improvements of the force models.

\section{Force estimation from particle-resolved fluid data}
\label{sec:prforces}
\subsection{Particle-resolved simulations}
\label{sec:prsims}

\begin{figure}
	\centerline{\includegraphics[]{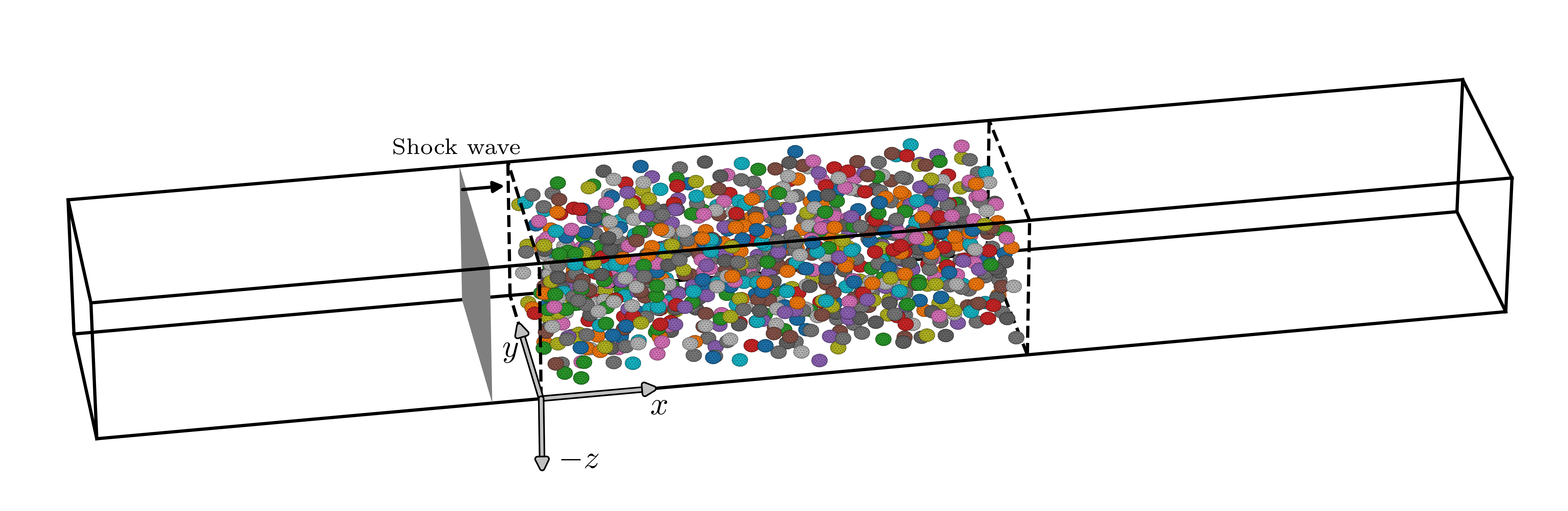}}
	\caption{Sketch of the problem setup and the computational domain. A particle layer with particle volume fraction 0.1 is located between $0\leq x \leq L$. A shock wave with Mach number 2.6 is initially located at $-0.1 L$ and propagating towards the particle layer. The particle Reynolds number based on post-incident shock values is 2000.}
	\label{fig:sketch}
\end{figure}

The particle-resolved simulations that are used as reference data in this work were described and analyzed in \citet{osnes2019c}. For the readers convenience, we repeat the most important details here. The problem considered is that of the propagation of an initially Mach 2.6 shock wave through a random array of particles at 10\% volume fraction. The simulations are three-dimensional and include viscous terms, and the particles are assumed to be stationary and inert. The particle Reynolds number is 
\begin{equation}
\Rep_{,\is}=\frac{\rho_\is u_\is \Dp}{\mu_\is} = 2000,
\end{equation}
where subscript $\is$ denotes a flow variable behind the incident shock wave, $\rho$ is the fluid density, $u$ is the fluid velocity in the downstream direction, $\Dp$ is the particle diameter, and $\mu$ is the fluid viscosity. \cref{fig:sketch} shows a sketch of the problem setup. The particle layer is located between $0<x<L$, where $x$ is the streamwise coordinate and $L$ is the particle layer length, which is $12\sqrt[3]{16}\Dp$. The spanwise extents of the domain are $8\sqrt[3]{4}\Dp$. The spanwise domain boundaries are periodic, while the upstream boundary is set to the post-shock state and the downstream boundary is a zero-gradient outlet. Ten simulations with different random distribution realizations were performed, and the results were volume-averaged as well as averaged over this simulation ensemble. The initial position of the shock wave was set to $-0.1L$, and each simulation was run until $t=3.75\tau_L$, where
\begin{equation}
\tau_L=\frac{L}{\Ma_\is c_0},
\end{equation}
is the time it takes for the initial shock wave to travel a distance equal to the particle layer length. Here, $c_0$ is the ambient speed of sound. The fluid equation of state is set to an ideal gas with adiabatic index $\gamma=1.4$.

The computational mesh consists of an unstructured Voronoi-based grid with a control volume length scale of approximately 0.036$\Dp$ for regions closer than $\Dp/2$ to any particle. In the rest of the domain, the control volume length scale is doubled, except from part of the upstream and downstream regions, where it is doubled again. The total number of control volumes is approximately $10^8$. This resolution was able to accurately capture the drag on an isolated particle and the viscous length scales were reasonably resolved at $\Rep_{,\is} \approx 5000$ \citep{osnes2019}. The simulations were also close to converged in terms of the velocity fluctuation levels \citep{osnes2019b}. Since the current study considers $\Rep_{,\is} = 2000$, the grid-requirements are less strict, and thus the flow field is better resolved in the current study than that for which the grid-resolution was tested. Additionally, we conduct two single-sphere simulations here in order to verify the ability of the particle-resolved simulations to capture the particle forces. These simulations are presented in \cref{sec:singlespheresimulations}. For further details about the the governing equations, the computational method, and other details about the simulations, the reader is referred to \citet{osnes2019c}. 

The particle-resolved simulation data contains the force histories for 9310 particles. The flow field data was recorded by volume-averaging over bins with a streamwise extent of $L/60\approx0.5\Dp$ and spanning the $y$ and $z$ directions. The averaged flow field is thus only a function of the streamwise spatial coordinate $x$ and time. All terms in the volume-averaged flow equations (see \citet{osnes2019d}) were stored. The volume-averaged results thus contain the data that would be available in an EL or EE simulation with a streamwise grid-spacing equal to the bin-width. In addition, the particle-resolved results include terms such as the correlations between velocity fluctuations, the correlation between pressure-fluctuations and velocity fluctuation, etc., that cannot be directly computed in an EL simulation. These additional terms give information about the disturbance fields induced by the particles. In \citet{osnes2019,osnes2019c}, a simple model for the average velocity disturbance field (or velocity component of the pseudo-turbulent fluctuations) was proposed. This model accounts for the fluctuations that are due to particle wakes, and was calibrated using the velocity fluctuation correlations. This model will be used below to approximate the undisturbed flow field, which is needed in the drag force models.  

\subsection{Particle force models}
\label{sec:forcemodels}
\begin{figure}
	\centering
	\includegraphics{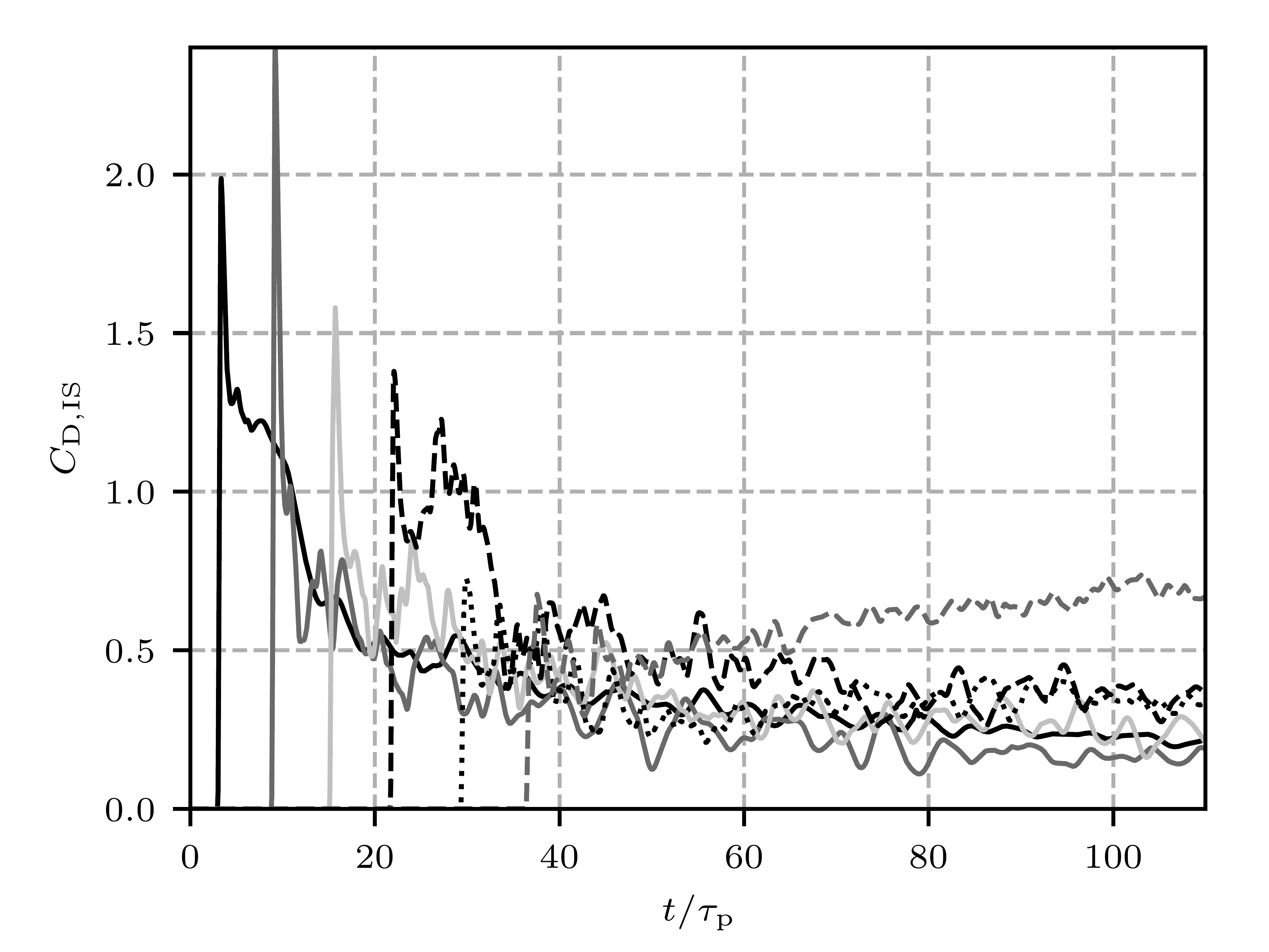}
	\caption{Example force histories at different locations. These are taken from the particles in the ensemble that have streamwise coordinates closest to the six locations $x/L=0,\ 0.2,\ 0.4,\ 0.6,\ 0.8,\ 1$.}
	\label{fig:force-example}
\end{figure}

Examples of the force histories that are to be approximated by the drag model is shown in \cref{fig:force-example}. The figure shows the drag coefficient in the streamwise direction for the six particles in the ensemble that are the closest to the six streamwise locations $x/L=0,\ 0.2,\ 0.4,\ 0.6,\ 0.8,$ and $1$, as a function of time. Here, time is normalized by the time it takes for the incident shock wave to travel one particle diameter, labeled $\tau_\mathrm{p}$ and given by
\begin{equation}
\tau_\mathrm{p}=\frac{\Dp}{\Ma_\is c_0},
\end{equation}
while the drag coefficient is given by
\begin{equation}
C_\mathrm{D,IS}=\frac{\int_{S_\mathrm{p}}\left(pn_1-\sigma_{1j}n_j\right)dS}{0.5\rho_\is u_\is^2\frac{\pi}{4}\Dp^2},
\end{equation}
where $S_\mathrm{p}$ is the particle surface, $p$ is the pressure, $n$ is the normal vector at the particle surface, repeated subscripts imply summation over components 1-3, where component 1 is in the streamwise direction, and $dS$ is a surface element.  For each particle, the force history starts with a very sharp peak when the shock wave impacts on the particle. Subsequently, the force decays rapidly as the shock wave moves further downstream. After about $10-20\tau_p$, the force attains a more stable value, but there are still significant variations about the slowly varying mean force. 

In \citet{parmar2012}, the following force decomposition is used to split the forces into different physical effects, 
\begin{equation}
F_\mathrm{p}=F_\mathrm{qs} + F_\mathrm{un} + F_\mathrm{iu} + F_\mathrm{vu},
\label{eq:force_decomposition}
\end{equation}
where $F_\mathrm{p}$ is the total force acting on the particle, while the four components are the quasi-steady drag, the undisturbed fluid (stress divergence) force, the inviscid unsteady force, and the viscous unsteady force. \citet{annamalai2017} derived the expressions for these based on a generalized Fax\'en's theorem. These are 
\begin{equation}
F_\mathrm{qs}=3\Dp\pi\nu\overline{\left(\rho u\right)\supun}^S\Phi(\Rep,\Map,\alpha_\mathrm{p}),
\end{equation}
where $\nu$ is the kinematic viscosity of the fluid, overline denote averaging over the particle surface (superscript S) or volume (superscript V), superscript "un" denotes undisturbed fluid properties, $\Phi$ is a drag correlation factor, typically given by an empirical correlation, and $\Map=(\overline{u}^V-u_\mathrm{p})/\overline{c}^V$ is the particle Mach number, where $u_\mathrm{p}$ is the particle velocity and $c$ is the local speed of sound, and $\alpha_\mathrm{p}$ is the particle volume fraction. Quasi-steady drag correlations are often given in terms of the drag coefficient, which is a function of far-field flow properties. In the case of a uniform flow field, the relation between $\Phi$ and the drag coefficient is
\begin{equation}
\Phi=\frac{\Rep}{24} C_\mathrm{D,qs}.
\end{equation}
The undisturbed fluid force is given by
\begin{equation}
F_\mathrm{un} = \Vp\overline{\rho\supun\frac{Du}{Dt}\supun}^V,
\end{equation} 
where $\Vp$ is the particle volume and $D/Dt$ is the material derivative based on the undisturbed fluid velocity. The inviscid unsteady force is
\begin{equation}
F_\mathrm{iu} = \Vp\int_{\xi=-\infty}^{t}K_\mathrm{iu}(t-\xi,\Map)\left(t-\xi\right)\left[\overline{\frac{D}{D t}}\overline{\left(\rho u_r\right)\supun}^S\right]_{t=\xi}d\xi,
\end{equation} 
where $K_\mathrm{iu}$ is the inviscid unsteady kernel, while the viscous unsteady forces are 
\begin{equation}
\begin{split}
F_\mathrm{vu} &= 18\Rp^2\sqrt{\pi\nu} \int_{\xi=-\infty}^{t}K_\mathrm{vu}^V\left(t-\xi,\Rep,\Map\right)\left.\frac{d}{d t}\overline{\left(\rho u_r\right)\supun}^S\right|_{t=\xi}d\xi\\
&+ 6\Rp^2\sqrt{\pi\nu} \int_{\xi=-\infty}^{t}K_\mathrm{vu}^S\left(t-\xi,\Rep,\Map\right)\left.\frac{d}{d t}\overline{\left(\rho u\right)\supun}^S\right|_{t=\xi}d\xi
\end{split},
\end{equation}
where $K_\mathrm{vu}^S$ is the viscous unsteady kernel for the surface contribution of the viscous unsteady force, $K_\mathrm{vu}^V$ is the kernel related to the volume contribution, $u_r$ is the radial component the velocity as observed by the particle, and $\frac{d}{dt}$ is the material derivative following the particle. In these expressions, the flow properties in $\Rep$ and $\Map$ should be taken as the averages of the undisturbed fluid properties over the particle volume. 

We use the following models for the different force components. For the quasi-steady drag, we use the model by \citet{parmar2010}, where 
\begin{equation}
C_\mathrm{D,qs}(\Rep, \Map) = \left\{\begin{tabular}{lll}
$C_\mathrm{D, std}(\Rep)+\left[C_\mathrm{D,M_\mathrm{cr}}(\Rep)-C_\mathrm{D,std}(\Rep)\right]\frac{\Map}{M_\mathrm{cr}}$& if & $\Map\leq M_\mathrm{cr}$\\
$C_\mathrm{D,sub}(\Rep,\Map)$ & if & $M_\mathrm{cr} \leq \Map \leq 1$\\
$C_\mathrm{D,sup}(\Rep,\Map)$ & if & $1 < \Map \leq 1.75$
\end{tabular}\right. .
\label{eq:parmarFqs}
\end{equation}
The reader is referred to \cite{parmar2009} for the expressions for the various parameters in \cref{eq:parmarFqs}. The undisturbed fluid force is predominantly the pressure gradient force, and will thus be computed as such. However, it is important to notice that it is the undisturbed flow properties that enter in this force as well.

The inviscid unsteady force is modeled with the Mach-number dependent kernels obtained by \citet{parmar2008,parmar2009} in tabulated form. It should be noted that the tabulated kernels are based on constant acceleration of a sphere in a fluid at a given background Mach number, where the fluid acceleration magnitude is significantly less than that of the incident high Mach number shock wave considered here. Therefore, it is not obvious that the kernel captures the relevant flow physics in the present configuration. Indeed, the results of \citet{parmar2009} indicate that the accuracy of the force model is highly dependent on the shock wave Mach number (see Figure 6 in \cite{parmar2009}).  

For the viscous unsteady kernels, we use the model of \citet{mei1992}, where the surface and volume contributions are not given separately, but rather a single kernel is used. The viscous unsteady force is then
\begin{equation}
\Fvu=3\pi\mu\Dp\int_{\xi=-\infty}^{t}K_\mathrm{vu}(t,\xi)\frac{d}{dt}\overline{\left(\rho u\right)\supun}^Sd\xi,
\end{equation}
where
\begin{equation}
K_\mathrm{vu}(t,\xi)=\left\{\left[\frac{4\pi (t-\xi)\nu}{\Dp^2}\right]^{1/4}+\left[\frac{\pi\left|u(\xi)-u_\mathrm{p}(\xi)\right|^3}{\Dp\nu f_H^3(\Rep)}(t-\xi)^2\right]^{1/2}\right\}^{-2},
\end{equation}
where $f_H=0.75+0.105\Rep$. Like for the inviscid unsteady kernel, it should be noted that the viscous unsteady kernel is obtained in a different flow regime than appropriate for the current problem, with small oscillations of the inflow.

The above force models are derived for isolated particles, where the particle volume fraction is negligible and the undisturbed flow velocity can be easily estimated. Neither of these statements are true for the present case. To account for these differences, two model corrections are considered. For some of the results presented below, we will use the model proposed in \cite{osnes2019,osnes2019c} to approximate an undisturbed flow velocity at each particle location. The model introduces a correction factor to the volume-averaged velocity, which is a function of particle Reynolds number and particle volume fraction. The model approximates the undisturbed streamwise flow velocity by 
\begin{equation}
u\supun = \tilde{u}\frac{\alpha}{\alpha-\alphasep(\alphap, \Rep)},
\label{eq:ucorr}
\end{equation}
where $\tilde{u}$ is the Favre-averaged velocity, and $\alpha_\mathrm{sep}$ represents the volume fraction of separated flow in particle wakes. Here, the separation volume fraction will be modeled as
\begin{equation}
\alphasep(\alphap,\Rep) = \alphap C(\Rep),
\label{eq:alphasep}
\end{equation}
where \citet{osnes2019d} found $C(\Rep)\approx 1.5$ for the current flow configuration. It should be noted that it would be appropriate to introduce a time-dependency in \cref{eq:alphasep}, since particle wakes and fluctuations are not generated instantaneously after the shock wave passes over a particle. One possible approach to this time-dependency could be to model $\alphasep$ using a history integral over the relative flow velocity for all particles in the control volume. However, such a model has not yet been developed, and is outside the scope of the current work. Along with the correction to the undisturbed flow velocity, the model predicts a velocity fluctuation correlation given by 
\begin{equation}
\phavg{u''u''}=\favg{u}^2\frac{\alphasep}{\alpha-\alphasep},
\label{eq:R00}
\end{equation}
where $\phavg{}$ denotes phase-averaging. \Cref{eq:R00} will be used in the EL simulations below.  

The second correction model consists of volume fraction correction factors, originally developed by \citet{sangani1991}, and used in \citet{ling2012} and \citet{theofanous2017} for the quasi-steady drag forces, the inviscid unsteady forces, and the viscous unsteady forces. This approach will also be compared with the particle-resolved forces here. With this approach, we do not use \cref{eq:ucorr}. Instead, $\Fqs$, $\Fiu$, and $\Fvu$ are multiplied by the volume fraction correction factors 
\begin{equation}
\phi_\mathrm{qs}(\alphap)=\frac{1+2\alphap}{(1-\alphap)^3}, \quad \phi_\mathrm{iu}(\alphap)=1+2\alphap,\quad \phi_\mathrm{vu}(\alphap)=\frac{1}{1-\alphap},
\label{eq:vfcorr}
\end{equation}
respectively. These corrections are based on simulations of oscillatory flow in the linear regime, and their applicability to the present flow conditions is uncertain. In particular, it is doubtful that the initial shock-accelerated flow can be well represented, since there is no time to communicate the geometric information of nearby particles during the time it takes for the shock wave to interact with the particle. In fact, there are indications that a reduction, rather than an amplification, of the inviscid unsteady force is appropriate for shock-particle interaction in random particle arrays \citep{koneru2020}.  

In the following, we will take three different approaches for computing the particle forces. The first approach is to use the force models directly, as if each particle is isolated in a flow whose average properties are those of the volume averaged flow properties from the particle-resolved simulations. This approach will be referred to as the isolated particle model. Secondly, we will apply the same force models, but with the undisturbed fluid velocity modeled by \cref{eq:ucorr}. This will be referred to as the velocity-corrected model. Lastly, we will use the particle volume fraction correction models, \cref{eq:vfcorr}, instead of the velocity-corrected model. This will be referred to as the volume-fraction corrected model.  It is worth noting that while the volume fraction corrections scale $\Fqs$, $\Fiu$ and $\Fvu$ by constant factors for the current problem, the velocity-corrected model affects $\Fqs$, $\Fiu$ and $\Fvu$ in a non-linear, flow field dependent, manner. It will therefore be more or less effective at different times and in different regions. 

\subsection{Single sphere simulations}
\label{sec:singlespheresimulations}

In order to verify the ability of the particle-resolved simulations to accurately capture the particle forces, we simulate the interaction of an isolated particle with both a weak expansion fan and a shock wave. The inviscid expansion fan was previously considered in \cite{annamalai2017}, who applied both direct numerical simulations (DNS) and the Fax\'en force model to compute the particle force. The shock-particle interaction was studied by \cite{sun2005}, who presented results from both an experimental study and a numerical simulation. In both cases, the initial condition for the simulations consist of two constant states separated by a discontinuity. For the inviscid expansion, we use the following states
\begin{equation}
\rho_L = 1.2635\ \mathrm{kg/m^3},\ u_L = 0,\ p_L=107313\ \mathrm{Pa},
\end{equation} 
\begin{equation}
\rho_R = 1.2635\ \mathrm{kg/m^3},\ u_R = 0,\ p_R=102203\ \mathrm{Pa},
\end{equation}
where $\rho_L,\ u_L,\ p_L$ denote the density, velocity and pressure on the left side of the discontinuity, while  $\rho_R,\ u_R,\ p_R$ denote the corresponding values at the right side of the discontinuity. The discontinuity is located at $x=1.25\Dp$, and the particle at $x=0$. 

For the shock-particle interaction, the initial condition consists of the states
\begin{equation}
\rho_L = 1.6582\ \mathrm{kg/m^3},\ u_L = 114.47,\ p_L=159060\ \mathrm{Pa},
\end{equation} 
\begin{equation}
\rho_R = 1.2048\ \mathrm{kg/m^3},\ u_R = 0,\ p_R=101325\ \mathrm{Pa},
\end{equation}
and the particle diameter and gas viscosity are set by requiring $\Rep=4900$. For both cases, results for grids with similar resolution to that used in the full particle cloud simulations are presented. Grid convergence studies for similar configurations are found in \cite{osnes2019d}. 

\begin{figure}
	\centerline{
		\includegraphics{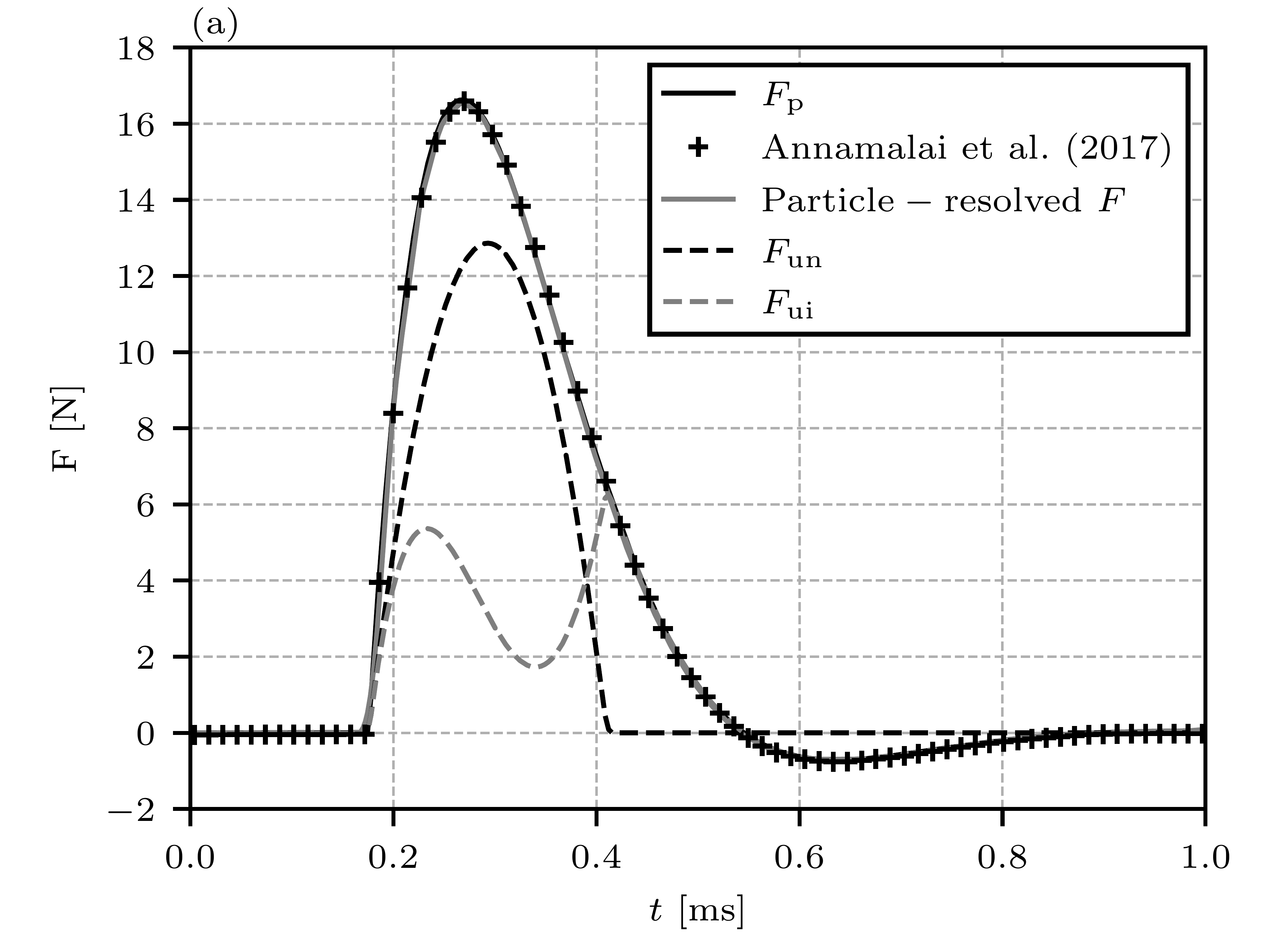}\includegraphics{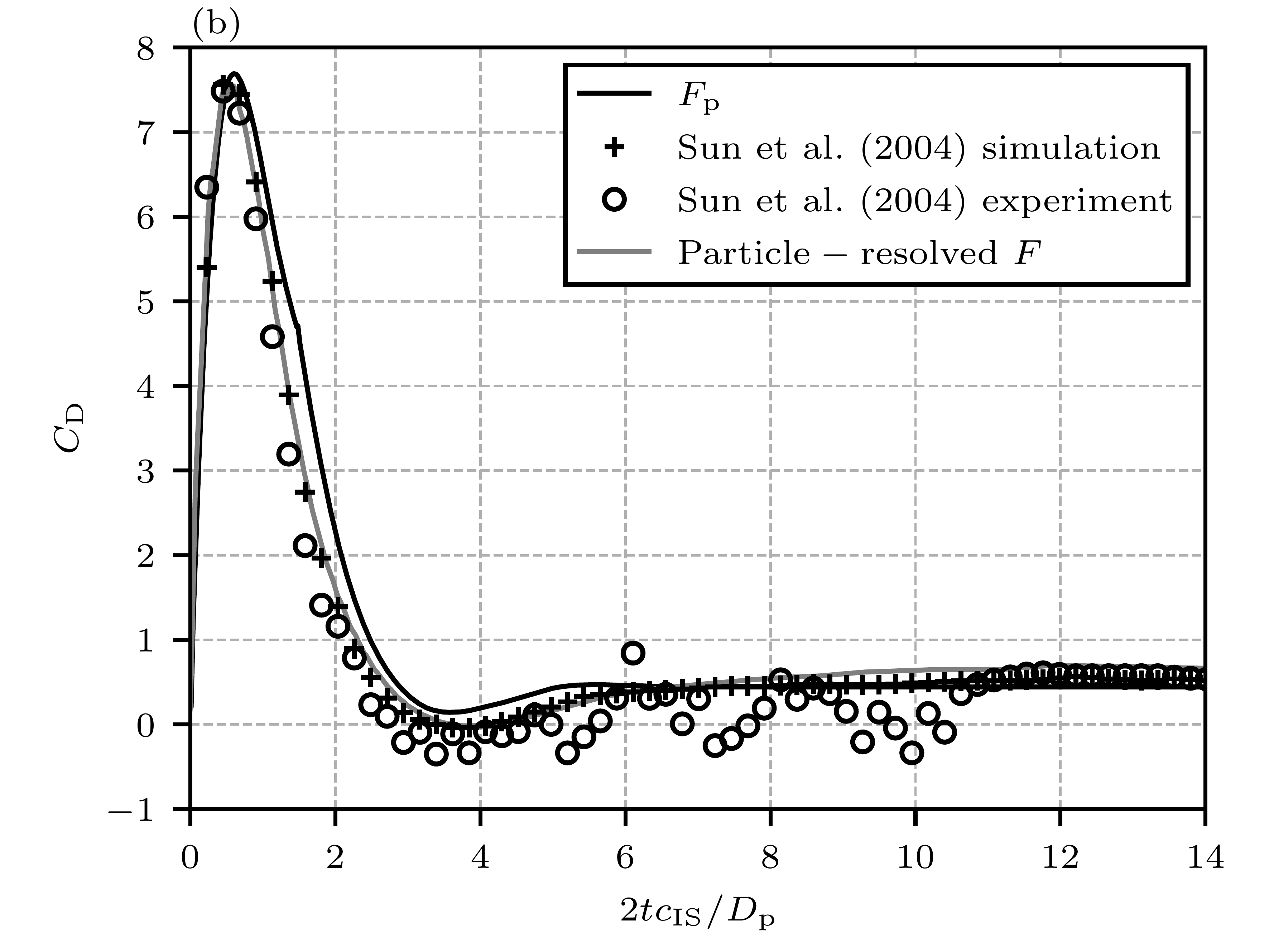}}
	\caption{Particle forces for (a): an isolated particle in a weak inviscid expansion fan, (b): an isolated particle exposed to a $\Ma=1.22$ shock wave at $\Rep=4900$. Here, $c_\mathrm{IS}$ denotes the speed of sound based on the post-shock state. }
	\label{fig:singleparticleforces}
\end{figure}

The simulation results are shown in \cref{fig:singleparticleforces} along with the (digitized) results of \cite{annamalai2017} and \cite{sun2005}, and the force model predictions. The force obtained in the particle-resolved simulation of the expansion fan agrees perfectly with the DNS results of \cite{annamalai2017}. This is also the case for the force model prediction, which, in this case, is based on the exact solution of the corresponding shock-tube problem without the particle. 

In the shock-particle case, the agreement of the current particle-resolved simulation and the simulations of \cite{sun2005} is excellent up to $2tc_\mathrm{IS}/\Dp=6$. At later times, the current simulation predicts a slightly higher force, which we attribute to the development of the particle wake, which can be expected to differ for axisymmetric and three-dimensional simulations. Compared to the experiments, slightly higher drag is obtained in the simulation. This is to be expected since the experiments were conducted at $\Rep\approx3\times10^5$, and thus the viscous force contribution is negligible in the experiments. The force model prediction is also in good agreement with the simulations, although not to the same extent as in the inviscid expansion case. Nevertheless, considering that the force comparisons presented in \cite{parmar2009} for shock-particle interactions also showed differences between the model prediction and the simulations of \cite{sun2005}, and that the viscous unsteady kernel is developed for a completely different flow regime, we consider the force model prediction shown in \cref{fig:singleparticleforces} to be quite good.

In conclusion, the particle-resolved simulations are able to accurately capture the forces on isolated particles. The force models also capture these forces very well. The results for forces in dense particle suspensions are presented next.

\subsection{Comparison of force models and particle resolved simulation results}
\label{sec:prforcecomparison}

\begin{figure}
	\centerline{\includegraphics[]{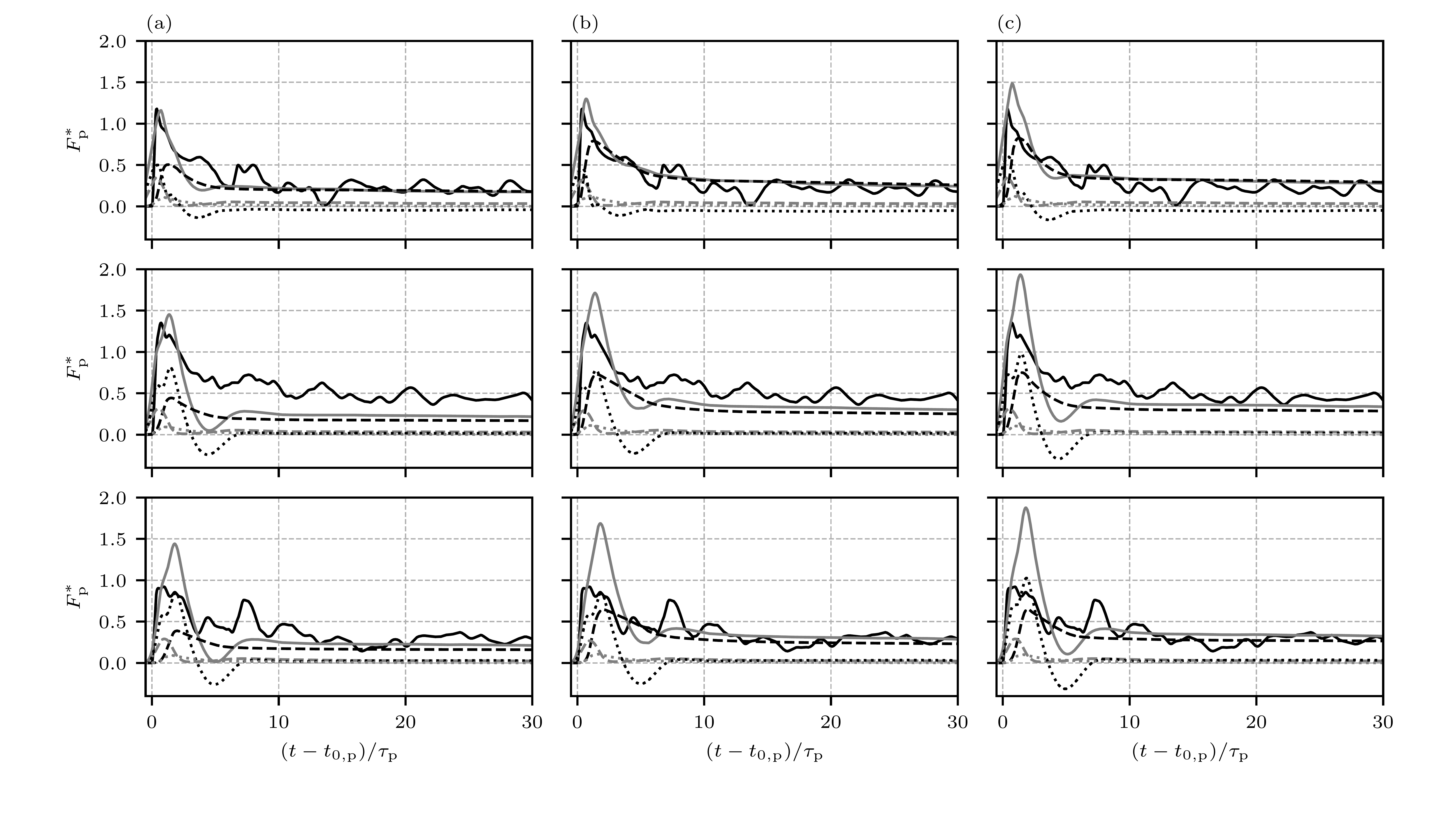}}
	\caption{Three particle histories as well as the drag force predictions. Solid black line: particle-resolved force, solid grey line: $F_\mathrm{qs}+F_\mathrm{un}+F_\mathrm{iu}+F_\mathrm{vu}$, dashed black line: $F_\mathrm{qs}$, dashed grey line: $F_\mathrm{un}$, dotted black line: $F_\mathrm{iu}$, dotted grey line: $F_\mathrm{vu}$. (a): isolated particle model, (b): velocity corrected model, (c): volume fraction corrected model.}
	\label{fig:force_and_basic_model_prediction}
\end{figure}

\begin{figure}
	\centerline{
		\includegraphics[]{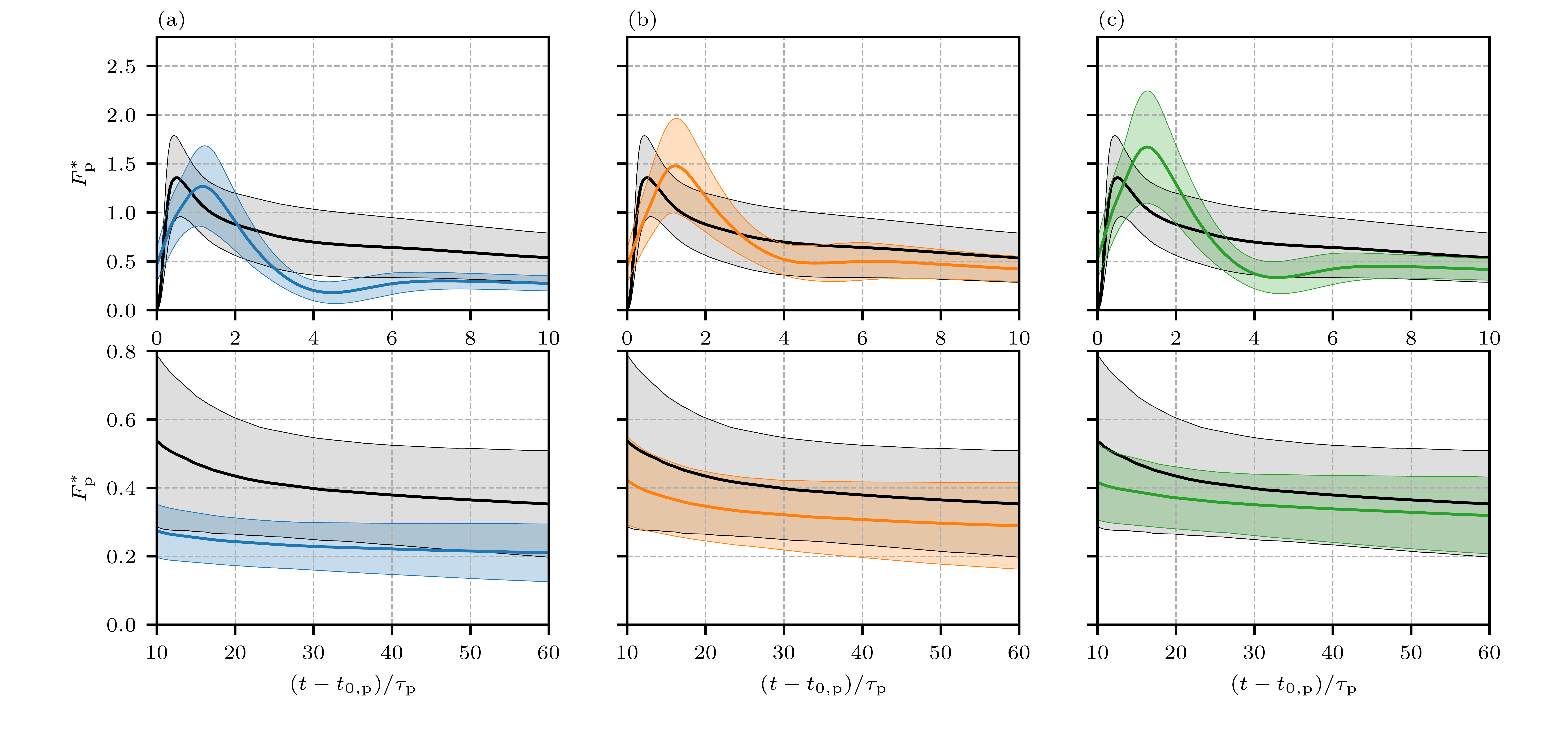}}
	\caption{Average forces. Black lines show the results for the particle-resolved simulations and the grey shaded regions show one standard deviation to each side. Colored lines show the average force model predictions, and the colored shaded regions show one standard deviation to each side. (a): Isolated particle model, (b): velocity corrected model, (c): volume fraction corrected model.}
	\label{fig:average_force_and_basic_model prediction}
\end{figure}

Individual particle histories are interesting to examine since these give an impression about the varied behaviour that the drag models should ideally predict. \Cref{fig:force_and_basic_model_prediction} shows three particle force histories along with the drag force predictions based on the models described above. Each particle history is shifted by a time $t_{0,\mathrm{p}}$, which is the time when the force on that particle first exceeds 1\textperthousand\  of the maximal particle force magnitude. The particles share a common general behaviour, with a sharp spike in the force, corresponding to the shock-particle interaction, followed by a rapid decay and then a more gradual slow decay over time. The rapid decay part is very different for these three particles, where especially the particle shown in the bottom panel has a behaviour that is significantly affected by fluid-mediated particle-particle interactions. Oscillations with a time-scale of about $2\taup-4\taup$ are clearly seen, and their amplitudes take values up to 50\% of the average drag, see e.g. the force around $(t-t_{0,\mathrm{p}})=14\taup$ for the particle shown in the upper panels. 

The initial spike in the force imposed by the shock wave, visible in the  histories, is predicted by the drag models. Its magnitude is overestimated for all three particles with all three approaches, except for the particle force shown in the top panel with the isolated particle model. The volume fraction corrected model predicts up to twice the magnitude observed in the particle resolved simulations, while the two other approaches have more modest overestimations. The primary factors responsible for the higher peaks in the volume-fraction corrected model is the strong amplifications of $\Fqs$ and $\Fiu$  in the volume fraction corrected model, since these peak at around the same time. This is in line with the comments made above, and consistent with the behaviour observed in Figure 6b of \citep{parmar2009} of scaling issues of the kernel with Mach number.

Some particle force histories are better predicted than others. For example, the particle force prediction in the upper-most panel for the isolated particle model fits better with the particle-resolved force than the one in the middle panel. This is expected due to the variation related to the particle configuration. This variation is not modeled with the drag laws evaluated here, and comparison of individual particle histories is therefore not a well-suited approach for evaluating the applicability of the drag laws to the current problem.  For this reason, we also consider the average predictions for the entire cloud.

Overall, none of the three approaches are able to capture the average drag forces very well. The most significant deficiencies are found in the average force magnitudes. \Cref{fig:average_force_and_basic_model prediction} shows the force history averaged over all particles, where the time is shifted by $t_{0,\mathrm{p}}$ for each particle, for the particle resolved simulations and the three modeling approaches. Qualitatively, the initial spike and subsequent decay are captured by all three approaches. The peak force is too high on average for the velocity corrected and the volume fraction corrected models, while it is too low for the isolated particle model. The peak force occurs slightly too late, but this is most likely related to the grid size used in this assessment.  At later times, the predicted mean values are lower than the particle-resolved mean values for all three approaches. The mean drag is approximately at one standard deviation below the particle-resolved mean force with the isolated particle model. With the velocity corrected model, this difference is more than halved, while the volume fraction corrected model is better again.

The force variations predicted by the drag models are smaller than the one from the particle-resolved simulations. The variation is largest for the volume-fraction corrected model and smallest for the isolated particle model. Underprediction of the force variation is expected, because the drag model will predict approximately the same force on all particles that have similar streamwise positions. The particle-resolved simulations also have a force variation that is related to the local particle configuration. Considering the entire particle cloud, slightly more than half the variation of the forces at most times can be explained by the differences in drag at different positions, and slightly less than half of the variation by the particle-configuration related drag differences.

\begin{figure}
	\centerline{
	\includegraphics[]{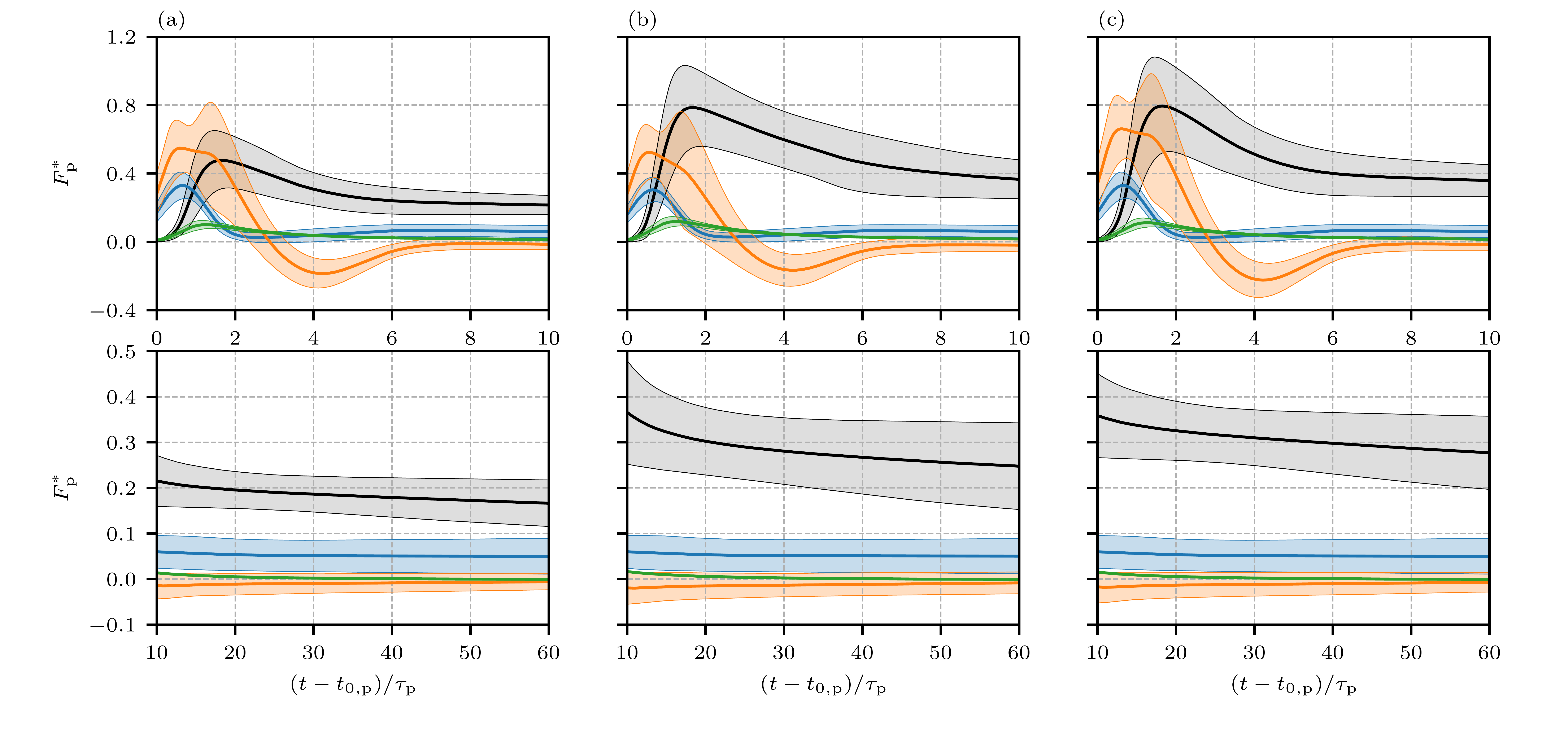}}
	\caption{Average force components. Black line: $\Fqs$, blue line: $\Fun$, orange line: $\Fiu$, green line: $\Fvu$. The shaded areas indicate one standard deviation. (a): Isolated particle model, (b): velocity corrected model, (c): volume fraction corrected model.}
	\label{fig:average_force_components}
\end{figure}

The various force components as a function of time are shown in \cref{fig:average_force_components}. The force has non-negligible contributions from all components in the decomposition, \cref{eq:force_decomposition}, but the unsteady forces are primarily important in a short interval after the shock impact. Over the time $t-t_{0,\mathrm{p}}\in [0,\taup]$, the inviscid forces ($\Fun$ and $\Fiu$) dominate. The magnitude of the inviscid unsteady force is significantly larger than the other components during this interval. It should be noted that the forces that involve gradients ($\Fun$ and $\Fiu$) will be slightly smoothed due to the coarse spatial sampling of the flow field relative to the thickness of the shock wave. The viscous contributions begin to be important at around $t-t_{0,\mathrm{p}}=\taup$. The viscous unsteady force has the smallest magnitude overall. The quasi-steady force dominates the drag after $t-t_{0,\mathrm{p}}=2\taup$. 

There are clear differences in the relative importance of the force components for the three approaches, and the scaling of the inviscid unsteady force appears to be inappropriate. In the volume-fraction corrected model, $\Fqs$, $\Fiu$ and $\Fvu$ are simply scaled up relative to the isolated particle model, giving them larger magnitudes relative to $\Fun$. First of all, considering that the peak drag force is higher than the particle-resolved results, this scaling appears to be inappropriate for the period of time following shortly after the shock wave passes over each particle. This is not entirely surprising, since the configuration information is unavailable on the timescale of the shock interaction, and the flow conditions are vastly different than for those where the corrections are derived. Secondly, at the later stages of the process, the drag predictions are too low and thus a scaling of positive forces appears to be appropriate. However, at this stage $\Fiu$ is negative, and therefore the scaling of this force acts in a way that brings the drag prediction further away from the level observed in the particle-resolved simulations. The velocity correction primarily affects the quasi-steady drag force, but it also has a small effect on the shape of the inviscid unsteady force history. Like the volume fraction correction, the velocity correction model was obtained for the quasi steady case, and direct application to the unsteady highly transient components does not appear to be appropriate.

The force models undershoot the average drag at around $(t-t_{0,\mathrm{p}})\approx2\taup-6\taup$, which is due to $\Fiu$. The inviscid unsteady force model produces a negative force contribution after $(t-t_{0,\mathrm{p}})\approx3$, resulting in underprediction of the total force. This negative force is predicted by the model because it is based on the flow around an isolated particle, where the negative force is a result of shock-wave diffraction and generation of a high-pressure region behind the particle \citep{sun2005}. However, in a random particle array, the particle positions are often such that many particles have other particles in their immediate proximity, which affects the shock-particle interaction through fluid-mediated particle-particle forces. This can potentially cancel out the negative inviscid forces. On average, it appears that the negative inviscid forces are significantly dampened by the particle distribution, since no local minimum is found in the average particle-resolved simulation results. This effect is likely to be a function of the average inter-particle distance and/or volume fraction, as as well as the local Mach number. The current results provide evidence that at 10\% volume fraction, the shock-diffraction patterns are sufficiently disrupted in a manner that removes the negative force on average. However, further studies are warranted in order to characterize how this depends on bulk flow properties. 

\begin{figure}
\centering
\includegraphics[]{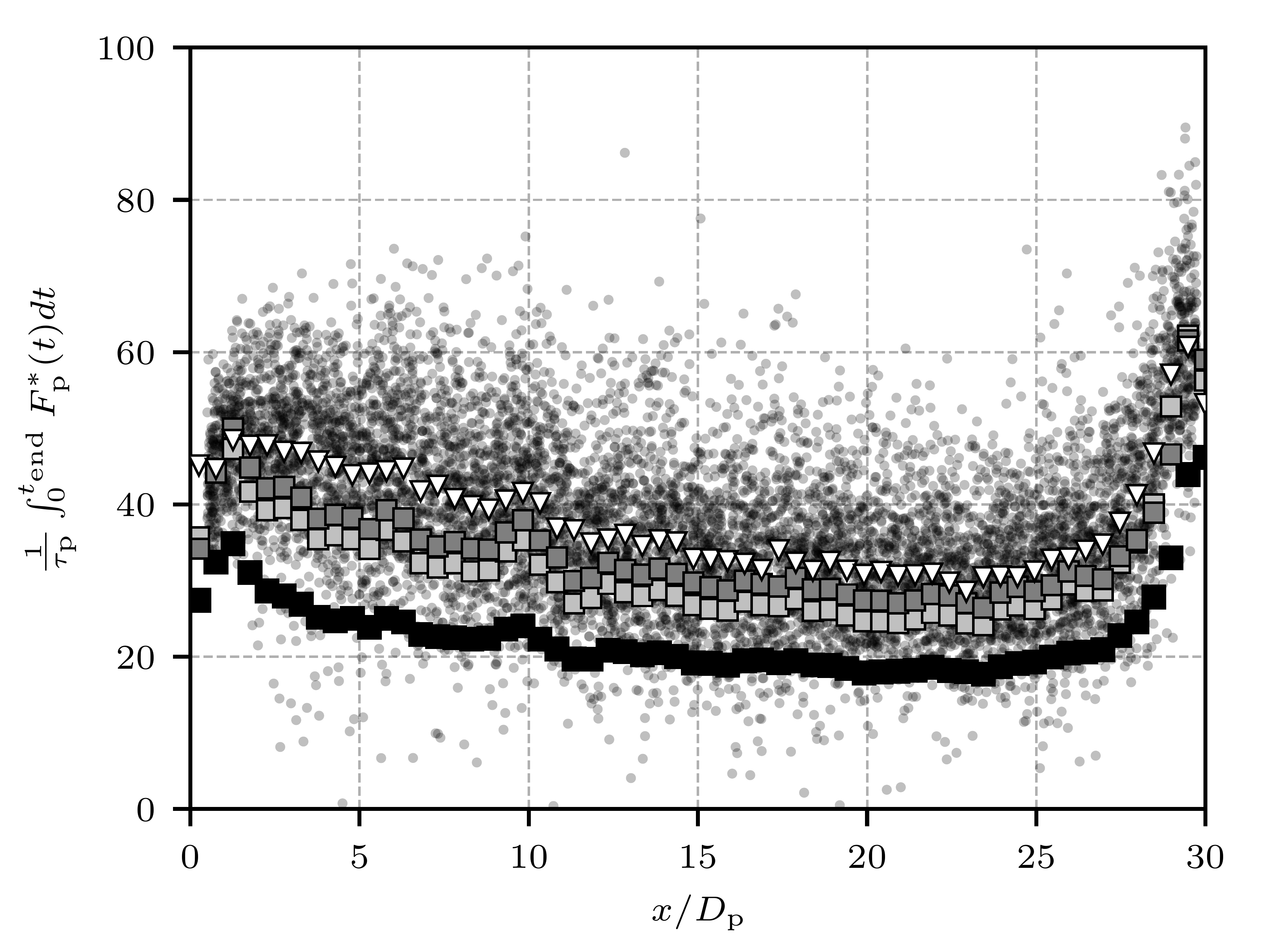}
\caption{Impulse for all particles over the full simulation time $t_\mathrm{end}\approx113\tau_\mathrm{p}$. Circles: particle-resolved data. Triangles: Spatially averaged particle-resolved data, black squares: spatially averaged isolated particle model,  light gray squares: spatially averaged velocity corrected model, dark grey squares: spatially averaged volume fraction corrected model.}
\label{fig:impulse_and_basic_model}
\end{figure}

For prediction of particle movement, the impulse is a better indicator than the force imposed at any specific time. Thus the impulse is a good quantity to consider when evaluating the overall performance of the drag laws. With the particle-resolved simulation data as input, the volume-fraction corrected model gives the best prediction of impulse, except in the downstream edge region, where the velocity corrected model is better. In \cref{fig:impulse_and_basic_model}, the impulse over the whole simulation time is shown for each particle, along with the impulses predicted by the drag models. The particle-resolved data is shown for each individual particle as well as spatial average over bins with width $L/60$. The impulses predicted by the drag models are only shown as spatial averages. The isolated particle model underpredicts the average impulse by up to 50\%. The comparisons with the velocity corrected model and the volume fraction corrected models are better, but they still underpredict the impulse. The velocity-corrected model is better at the downstream edge due to the effect of increased Mach number caused by increased velocity. As shown in \cite{osnes2019}, the increased forces at the downstream edge in the particle-resolved simulations are due to a rapid increase in the Mach number. The drag forces depend strongly on Mach number in the transonic regime, see e.g. \citet{nagata2016}, and the drag law of \citet{parmar2010} accounts for this dependence. Thus, the velocity correction is more impactful around the downstream edge due to the higher local Mach numbers here. The drag law of \citet{parmar2010} captures the increased forces at the downstream edge, and this has been shown to be particularly important for simulations of shock-particle cloud interaction, since accumulation, rather than dispersion, of particles near the downstream edge has been shown to be a consistent problem in both EE and EL simulations \citep{theofanous2017}. Still, the consistent underprediction of impulse even with properly Mach number dependent drag laws indicates that simulations that use these drag laws are likely to underpredict particle movement and dispersion during the shock-particle cloud interaction.

\begin{figure}
\centerline{
\includegraphics[]{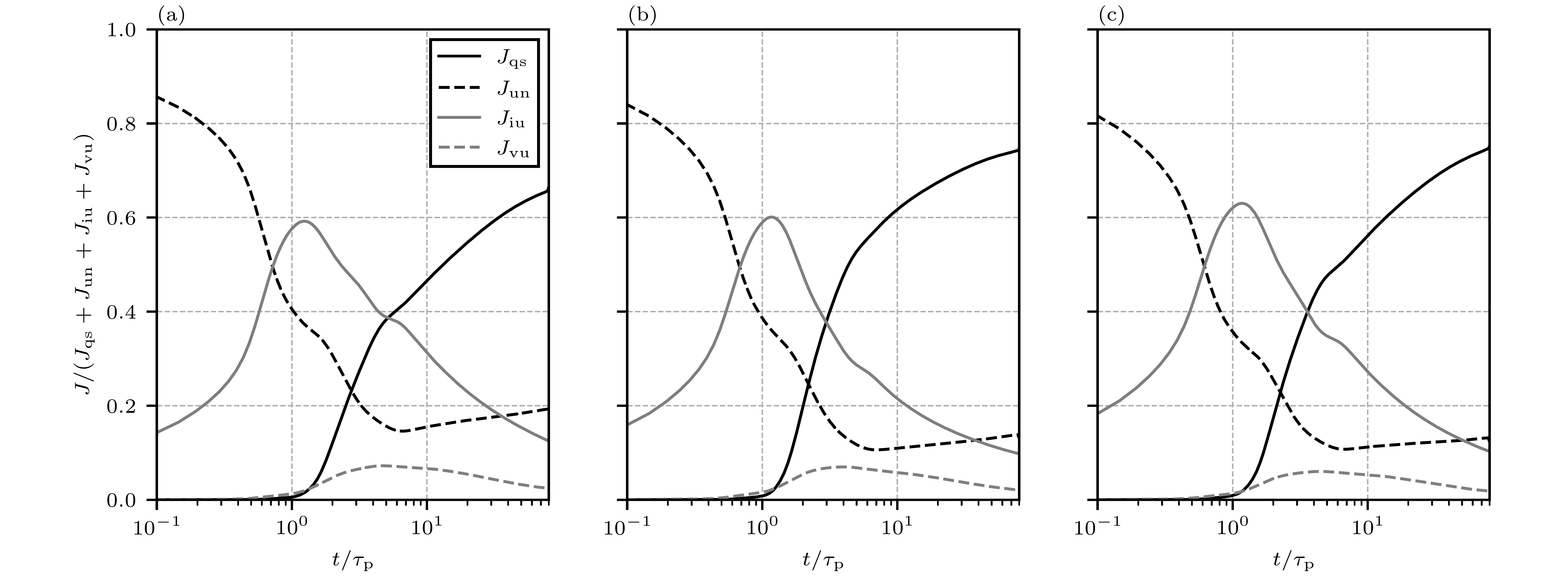}}
\caption{Impulse as a function of time for the different force components. (a): Isolated particle model. (b): velocity-corrected model. (c): volume-fraction corrected model}
\label{fig:impulse_components}
\end{figure}

The unsteady forces, and the corresponding impulses, are important only for a short time after the shock wave passes over each particle.  \Cref{fig:impulse_components} shows the impulse 
\begin{equation}
J = \int_0^{t}\sum_{i=0}^{N_\mathrm{p}}\left|F_\mathrm{p}(t-t_{0,\mathrm{p}})\right|
\end{equation}
for the different force components as a function of time for the three models. Note that the force magnitude is used to compute $J$, and thus $J$ is not directly translatable to particle momentum. The impulses are normalized by the sum of the four components, and thus the figure shows the relative importance of the forces in relation to particle motion, as a function of time. Early on, the undisturbed fluid force dominates. Around $t\approx0.5\taup$ the inviscid unsteady force is most important. Slightly later, the quasi-steady force impulse increases, and overtakes both the pressure-gradient impulse and the inviscid unsteady impulse before $t=4\taup$. The viscous unsteady force contributes by at most 9\% to the impulse, which occurs around $t\approx3\taup-5\taup$ (depending on the model). As time goes on, the impulse due to both unsteady forces becomes less and less important. The figure shows that only particles that have response time-scales of less than $\mathcal{O}\left(10\taup\right)$ will have a motion that is noticeably affected by the inviscid forces. However, the response of the gas to these forces can still be such that the inclusion of unsteady forces is important even for particles with large response time-scales.  

There are very small differences in the relative importance of the force terms for the three models. The most noticeable difference is that the pressure gradient impulse is more significant for the isolated particle-model when compared to the velocity-corrected and the volume-fraction corrected models at the later stages.

\section{Comparison of particle-resolved and EL simulations}
\label{sec:ELforces}
\subsection{EL simulation approach}
\label{sec:ELmethod}
The governing equations for the EL simulations are the volume-averaged mass, momentum and energy conservation equations. Since the problem under consideration is one-dimensional, only the corresponding one-dimensional volume-averaged equations will be given here. Additionally, the particles will be assumed identical, stationary, and inert. In the following, $\phavg{\psi}$ denotes a phase-averaged quantity, where $\psi$ is any fluid quantity, and is related to the volume-averaged value, $\vavg{\psi}$, through $\alpha\phavg{\psi}=\vavg{\psi}$. Additionally, $\favg{\psi}=\phavg{\rho\psi}/\phavg{\rho}$ denotes a Favre-averaged quantity. The deviation from the Favre-averaged quantity will be denoted $\psi''$. With these definitions, the one-dimensional volume-averaged conservation equations are

\begin{equation}
\pdts{\alpha\phrho} +\pdx{\alpha\phrho\fu} = 0,
\label{eq:vavgmass}
\end{equation}

\begin{equation}
\pdt{\alpha\rho\fu} + \pdx{\alpha \rho \fu^2 + \alpha \php} = \pdx{\alpha\frac{4}{3}\langle\mu\pdxs{u}\rangle}-\pdx{\alpha\phrho\wfavg{u''u''}}- \frac{1}{V}\sum_{i=0}^{N_\mathrm{p}}(F_\mathrm{qs,i}+F_\mathrm{un,i}+F_\mathrm{iu,i}+F_\mathrm{vu,i}),
\label{eq:vavgmom}
\end{equation}

\begin{equation}
\pdt{\alpha\phrho \tilde{E}} + \pdx{\alpha\phrho\tilde{E}\fuj+\alpha\php\fuj}=\pdx{\alpha\frac{4}{3}\langle\mu\pdxs{u}\rangle\fu} -\pdx{\alpha\phavg{\lambda\pdxs{T}}}- \pdx{\alpha\phrho\wfavg{u''u''}\fui} + \xi,
\label{eq:vavgE}
\end{equation}
where $\mu$ is the dynamic viscosity, $\favg{E}=\fe + \frac{1}{2}\fu^2+\frac{1}{2}\wfavg{u''u''}$ is the total energy per unit mass, where $e$ is the internal energy per unit mass, and $\lambda$ is the heat conductivity, which is assumed to be related to the viscosity through a constant Prandtl number of $0.7$. Note that the stationary particles do no work in the Eulerian frame, and thus there are no drag-related terms in the energy conservation equation. In the energy conservation equation, most of the sub-grid scale terms have been collected in the term $\xi$, which will be set to 0 in this work. The importance of these terms is not well known for shock-wave particle cloud interaction, and to the authors knowledge, no appropriate models exist for these terms under the current flow conditions. The omitted sub-grid scale terms can be found in \citet{osnes2019d}. \citet{mehta2020} quantified the inviscid sub-grid terms using inviscid particle resolved simulations, and found that both the internal energy-velocity correlation and the pressure-velocity correlation were negligible. The remaining sub-grid terms are the third velocity moment and terms involving the mass-weighted turbulent velocity, which have previously been modeled with an additional transport equation \cite{schwarzkopf2015}. The mass-weighted third velocity moment can be neglected based on the principle of receding influence. In \cite{vartdal2018}, the production of fluctuation kinetic energy by mass-weighted turbulent velocity terms was found to be small in comparison to the other production terms. Based on this, it can be assumed that the mass-weighted turbulent velocity diffusion terms in the total energy equation are also small, and thus they will be neglected here. An in-depth assessment of the importance of these terms is a topic for future studies.  

Only the volume-fraction corrected model will have a non-zero value for $\wfavg{u''u''}$, which is obtained by assuming that $\wfavg{u''u''}\approx\phavg{u''u''}$ and using \cref{eq:R00}. This means that the density fluctuation is not correlated with the magnitude of the velocity fluctuation, which is unlikely in general, but a more detailed model is necessary to account for this correlation. 

The equation of state for the gas is the ideal gas law, where internal energy, pressure and density are related by
\begin{equation}
p = (\gamma-1)\rho e,
\end{equation}
with $\gamma=1.4$. The temperature, $T$ is related to the internal energy by a constant heat capacity, and we assume that the viscosity varies with temperature as 
\begin{equation}
\mu(T) = \mu_\mathrm{ref}\left(\frac{T}{T_\mathrm{ref}}\right)^{0.76},
\end{equation}
where $\mu_\mathrm{ref}$ is the value of the viscosity at the reference temperature $T_\mathrm{ref}$. 

The conservation equations \cref{eq:vavgmass,eq:vavgmom,eq:vavgE} are solved with a control-volume based finite volume method. The fluxes between control volumes are computed with a modified HLLC Riemann solver, and MUSCL reconstruction with the minmod limiter is used to define the Riemann problems. The solution is advanced in time with a third-order Runge-Kutta scheme. The fluid variables are interpolated linearly to the particle positions. Several interpolation points are used for each particle. These points are assigned weights depending on their position relative to the particle center, and the sum of the interpolated values to each of these points multiplied by the point's weight approximates the volume-average of the fluid property over the particle volume. The force from each particle on the fluid is transferred with a filter to ensure consistency upon mesh refinement. We use the two-step filtering approach of \citet{capecelatro2013}, where a mollification is first used to transfer the Lagrangian data to the Eulerian mesh, followed by a diffusion step to obtain the desired spatial distribution of the transferred terms. The filter width recommended in \citet{capecelatro2013} was $\sigma_\mathrm{f}=3\Dp$, which is what we use here. 

We use a four times finer spatial resolution in the EL simulations than the sampling-resolution of the volume-averaged particle-resolved data. The spatial resolution for the EL simulations is therefore approximately $\Delta x = 0.125\Dp$. The grid resolution does have an influence on the simulation results, and grid-converged results can not be expected at least until the filter kernels are well resolved. The dependence of the simulation results to grid resolution is shown in \labelcref{sec:appA}. Since the objective of the current work is to evaluate the applicability of the drag-laws, we accept spatial resolution that is likely not achievable in simulations of many real-world problems. These simulations will often be unable to resolve any flow at the particle scale. However, only a few features in the results are very sensitive to the grid resolution. These are the speed at which the reflected shock wave is generated, and the flow expansion at the downstream particle cloud edge.

\subsection{Simulation results}
\label{sec:ELsimulations}
\begin{figure}
\centerline{\includegraphics{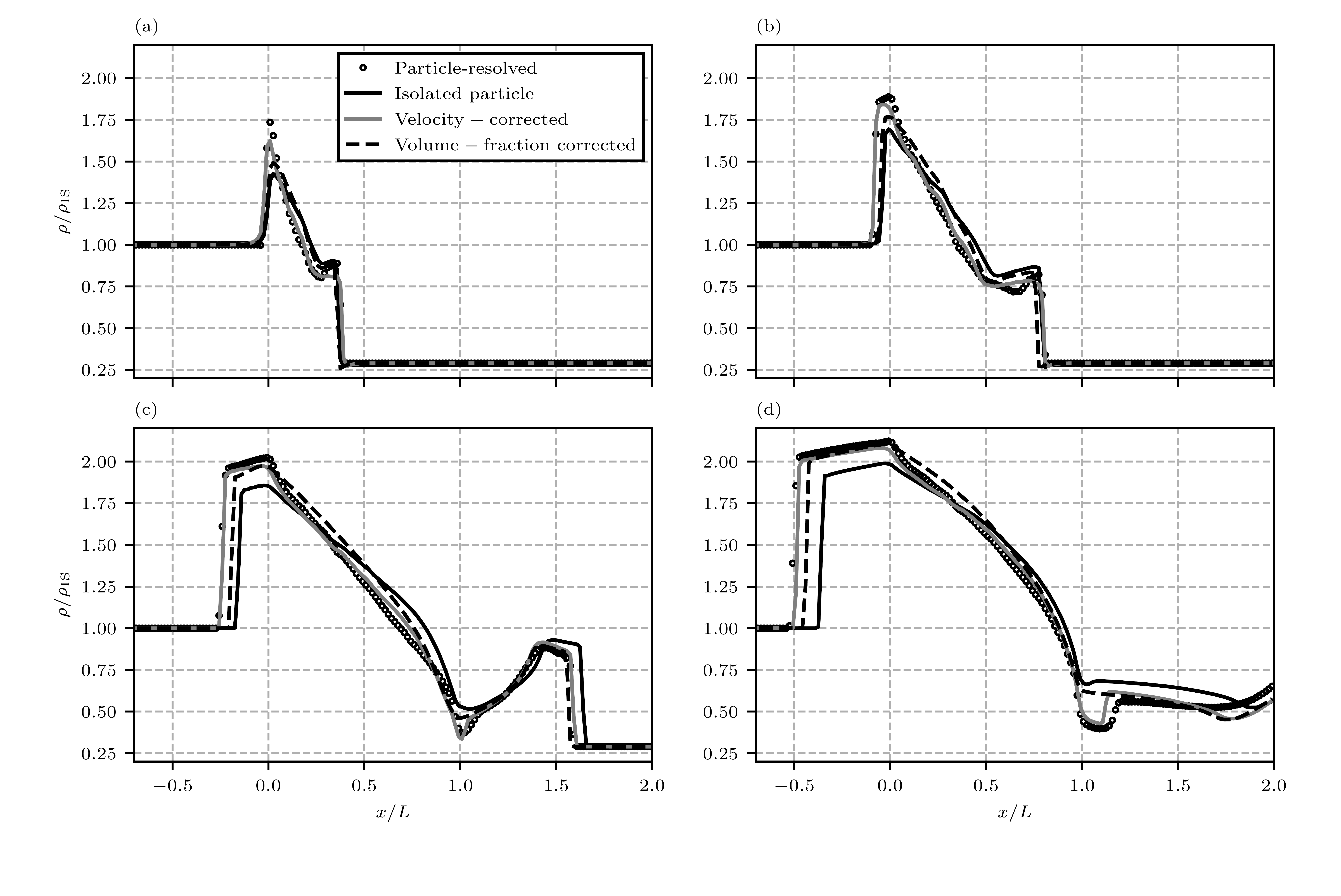}}
\caption{Density profiles at four times. (a): $t=15\taup$, (b): $t=30\taup$, (c): $t=60\taup$, (d): $t=100\taup$.}
\label{fig:EL_rho_results}
\end{figure}

In general, the mean flow fields are quite well captured with all three approaches, but the velocity corrected model is slightly better than the others. 
\Cref{fig:EL_rho_results} shows the mass density at four different times for the EL simulations and the particle resolved simulations. Early on, the EL simulations agree well with the particle-resolved simulations. The strength of the reflected shock wave is well predicted with the velocity-corrected model early on, while the two other models have a slightly weaker reflected shock. The volume-fraction corrected model gives a stronger reflected shock than the isolated particle model, as expected due to the stronger forces imposed on the flow by the particles. The stronger shock reflection with the velocity-corrected model can be attributed to the velocity-fluctuation term, which imposes an extra upstream-directed force on the flow at the upstream particle cloud edge. Similarly, an additional force appears at the downstream edge, this time directed downstream. This leads to a stronger flow expansion at the downstream edge, which agrees well with the one observed in the particle-resolved simulations. The isolated-particle model and the volume-fraction corrected model barely predict any overexpansion at all, and thus the agreement in the downstream region becomes poor at the later time points. 
It should be noted that the expansion region is particularly sensitive to the grid size, as shown in \labelcref{sec:appA}.

\begin{figure}
	\centerline{\includegraphics{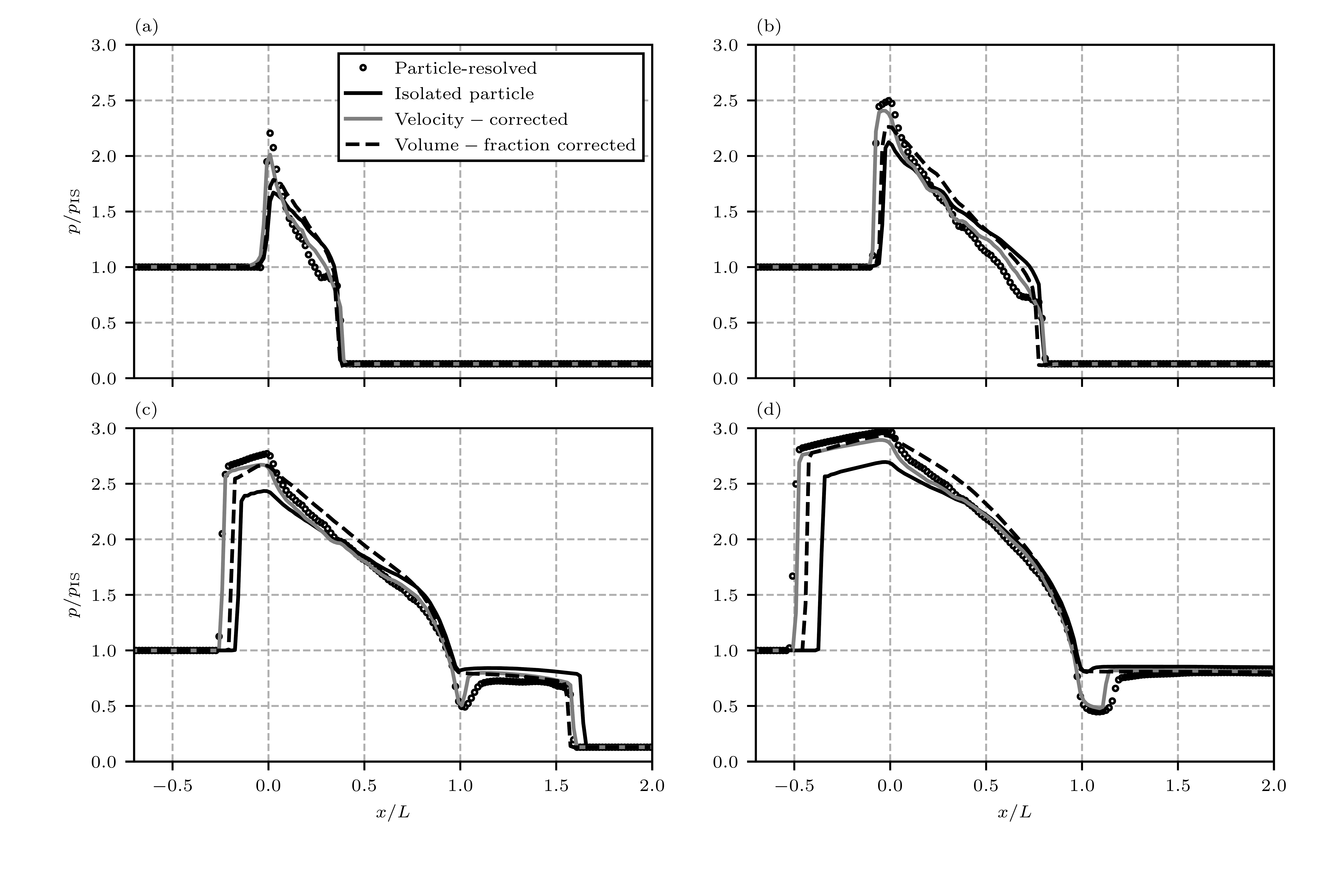}}
	\caption{Pressure profiles at four times. (a): $t=15\taup$, (b): $t=30\taup$, (c): $t=60\taup$, (d): $t=100\taup$.}
	\label{fig:EL_p_results}
\end{figure}

\begin{figure}
	\centerline{\includegraphics{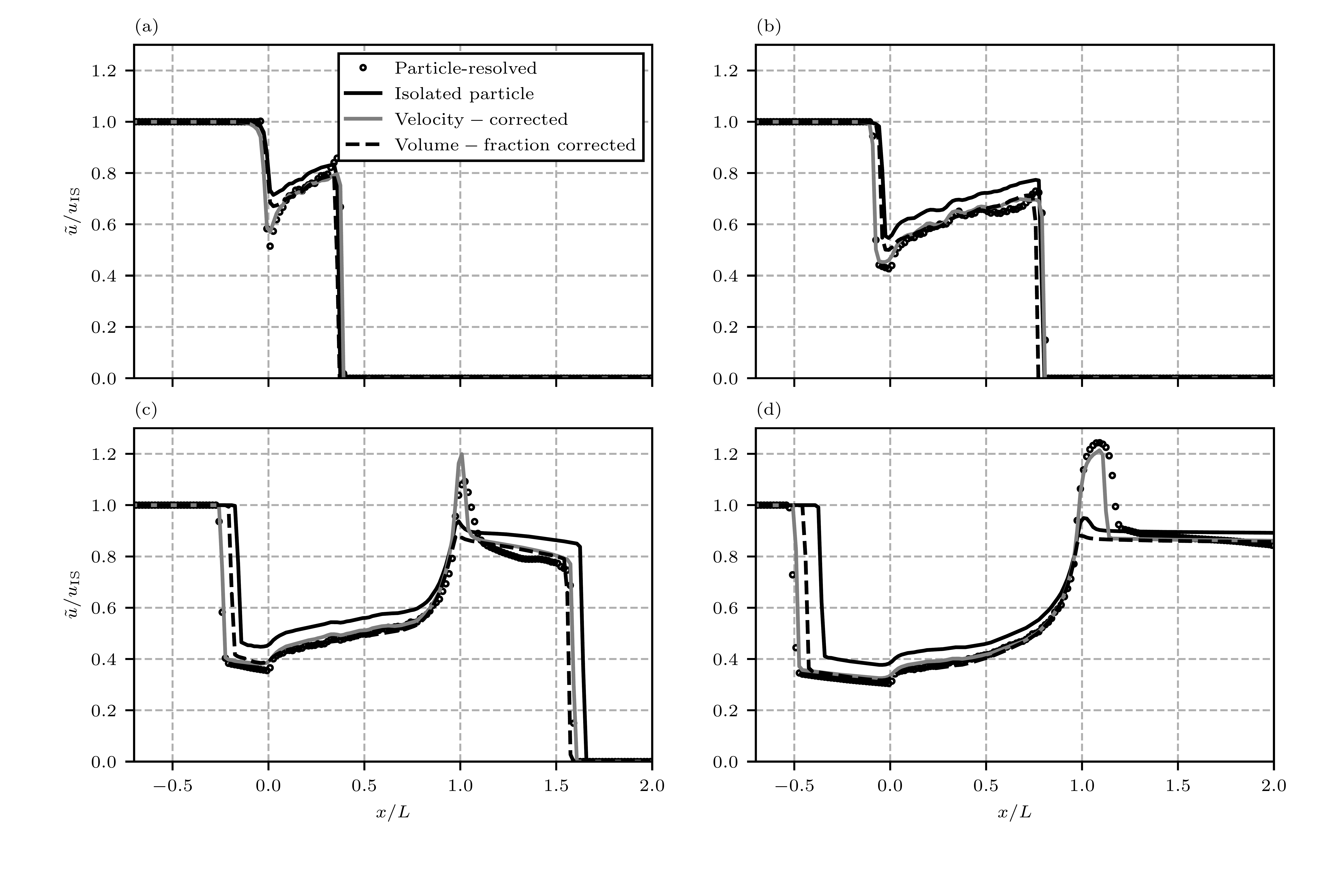}}
	\caption{Velocity profiles at four times. (a): $t=15\taup$, (b): $t=30\taup$, (c): $t=60\taup$, (d): $t=100\taup$. }
	\label{fig:EL_u_results}
\end{figure}

The corresponding pressure and velocity fields are shown in \cref{fig:EL_p_results,fig:EL_u_results}. The velocity results of the velocity corrected model is slightly higher than the particle-resolved velocity field in the expansion region at $t=60\taup$. In \citet{osnes2019,osnes2019c}, the authors found that the velocity fluctuations decreased rapidly at the downstream edge, but not to zero, and not as sharply as predicted by the velocity-corrected model. The too sharp velocity fluctuation decay predicted by the model produces a strong streamwise force that appears immediately after the shock wave passes. In the particle-resolved simulations, there is a slight time-delay before significant velocity fluctuations appear, and it is therefore not surprising that the flow expansion is stronger early on with the velocity corrected model.

\begin{figure}
	\centerline{
		\includegraphics[]{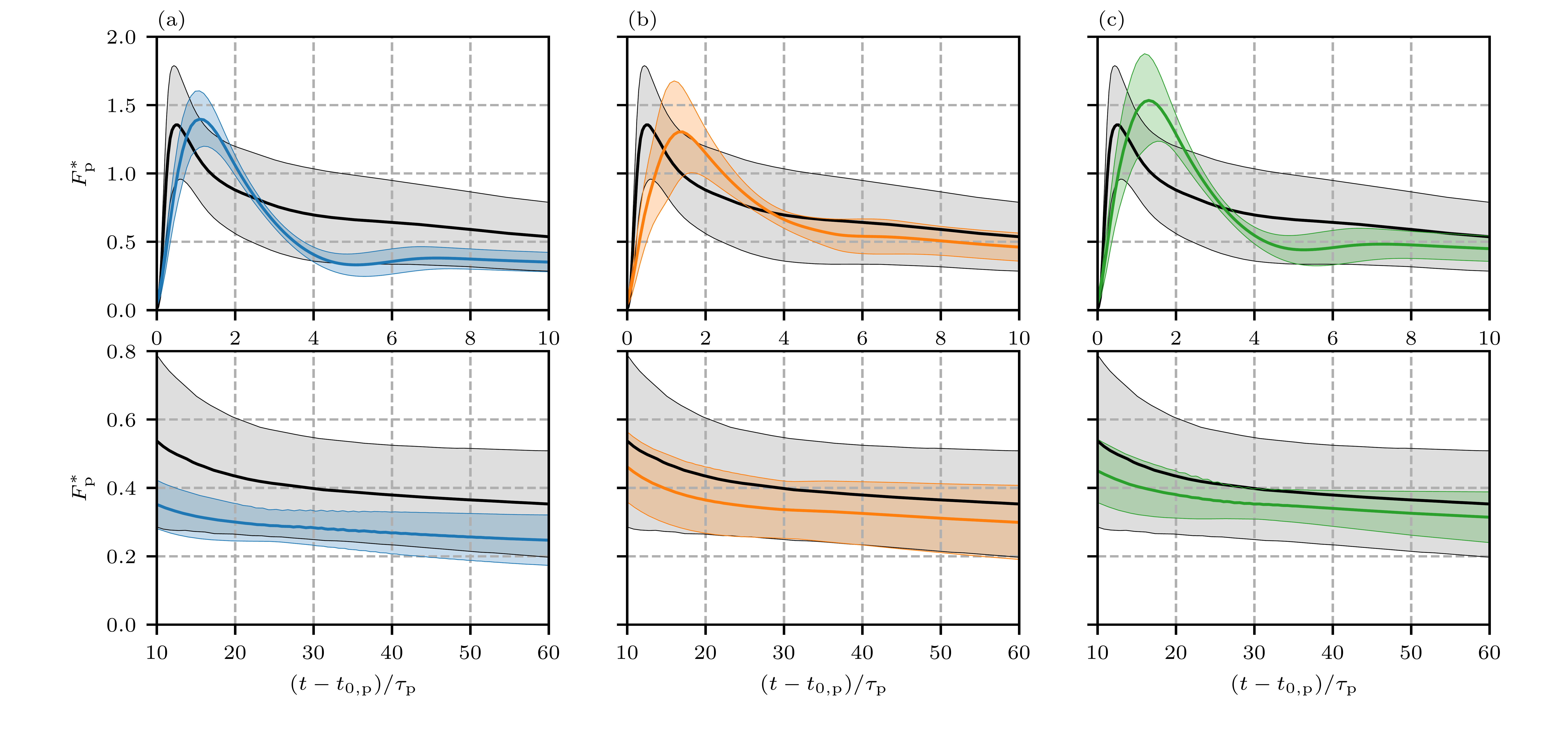}}
	\caption{Average particle force predictions for all particles in the particle-resolved simulations (black lines), and for the EL simulations (colored lines). The shaded areas indicate one standard deviation. (a): Isolated particle model, (b): velocity corrected model, (c): volume fraction corrected model.}
	\label{fig:EL-average-forces}
\end{figure}

\begin{figure}
	\centerline{
		\includegraphics[]{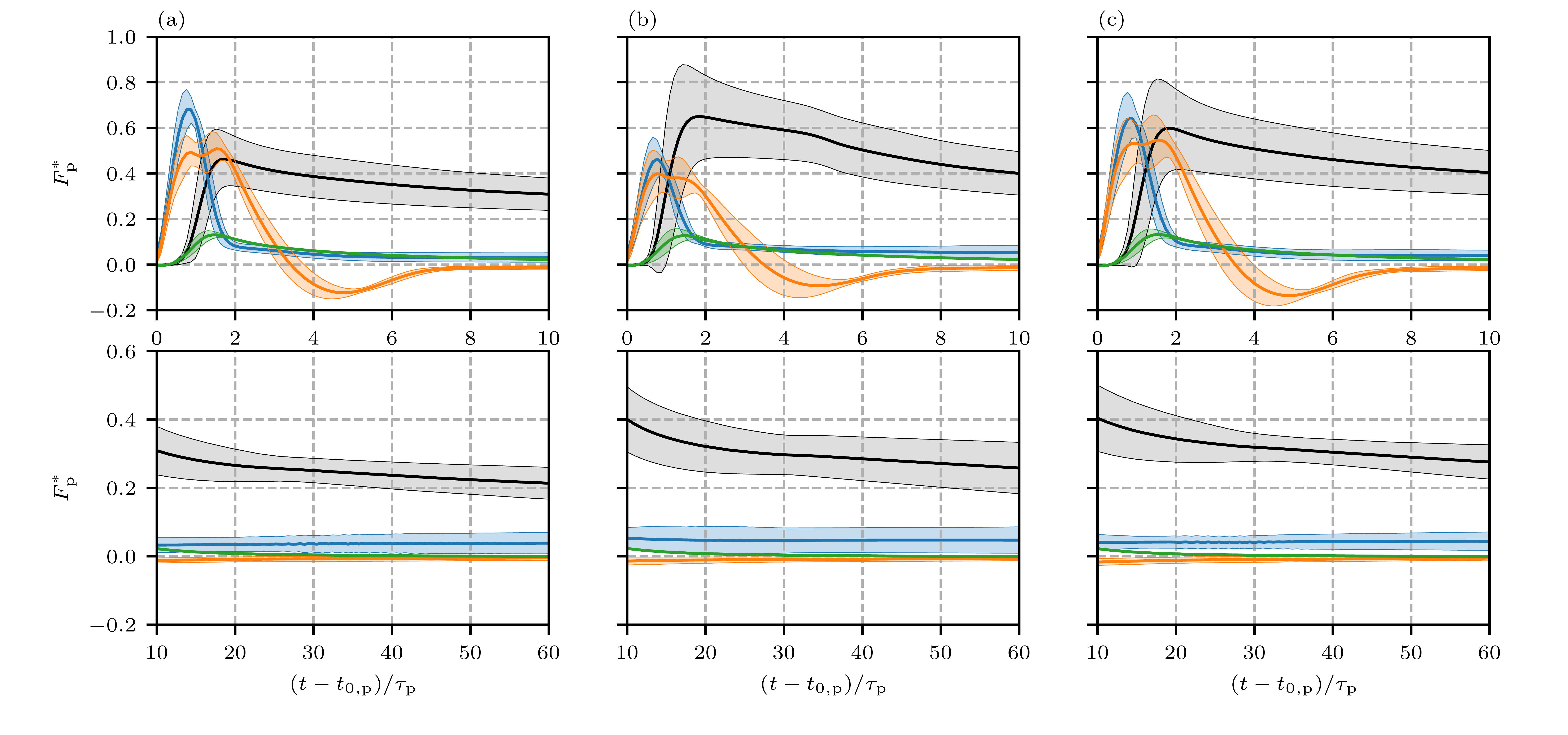}}
	\caption{Average force components over time in the EL simulations, for the three models. Black line: $\Fqs$, blue line: $\Fun$, orange line: $\Fiu$, green line: $\Fvu$. The shaded areas indicate one standard deviation. (a): Isolated particle model, (b): velocity corrected model, (c): volume fraction corrected model.}
	\label{fig:EL-average-force-components}
\end{figure}

\Cref{fig:EL-average-forces} shows the average particle forces in the EL simulations along with the averaged forces in the particle-resolved simulations. \Cref{fig:EL-average-force-components} shows the contribution of the different force components in the EL simulations. In these simulations, like in the case where we used volume-averaged particle-resolved data as input, we find that the peak force is overpredicted by the volume fraction corrected model. Here, this is also the case for the isolated particle model, but the velocity corrected model underpredicts it instead. The time of the peak force is also slightly delayed compared to the one in the particle-resolved simulations. However, since the force history for each particle is shifted with $t_{0,\mathrm{p}}$, the definition of this time-point has a direct impact on the location of the peak value, especially when the computational grid is not very fine, so that the numerical smoothing of the flow fields leads to an earlier $t_{0,\mathrm{p}}$ for the EL simulations. Nevertheless, the shock related transient is significantly smoothed in time, and the forces do not increase as quickly following shock impact as for the particle-resolved simulations. Finer Eulerian grids will have an effect on this, and will likely lead to more rapid force increases. However, this is also likely to increase the peak value further, and thus bring the peak forces further away from those observed in the particle-resolved simulations. The volume-fraction corrected force is still the highest, but the difference is smaller than it was in the results presented in the previous section.  

At late time, the forces with the volume-fraction corrected model are again those that are closest to the particle-resolved simulations, but they are in less agreement than when the forces were evaluated with the volume-averaged particle-resolved flow fields. This is related to the differences in the mean flow fields, c.f. \cref{fig:EL_rho_results,fig:EL_p_results,fig:EL_u_results}. The velocity-corrected model results are similar at late times as they were with the volume-averaged particle-resolved data as input, while the results of the isolated particle model are surprisingly slightly better than they were previously. 

Considering the force components, the largest differences between the three models are found in the quasi-steady forces and the undisturbed flow force. The quasi-steady force  is largest for the velocity-corrected model early on, but is largest for the volume-fraction corrected model at late time. It is significantly lower for the isolated particle model at all times. The undisturbed flow force is lower early on for the velocity-corrected model than the isolated particle model and the volume-fraction corrected model. For the isolated particle model, this can be explained by the weaker shock wave attenuation which leads to a stronger shock wave and therefore stronger pressure forces. For the volume-fraction corrected model, the increased undisturbed fluid force is more surprising, since it is the only force component that is not increased by a scaling factor. It is a result of an increased pressure gradient, which is induced by larger forces and no correction due to velocity fluctuations in the equation of state. The inviscid unsteady force is also slightly higher in the volume-fraction corrected model, which again can be attributed directly to the scaling factor.

\begin{figure}
	\centerline{
		\includegraphics[]{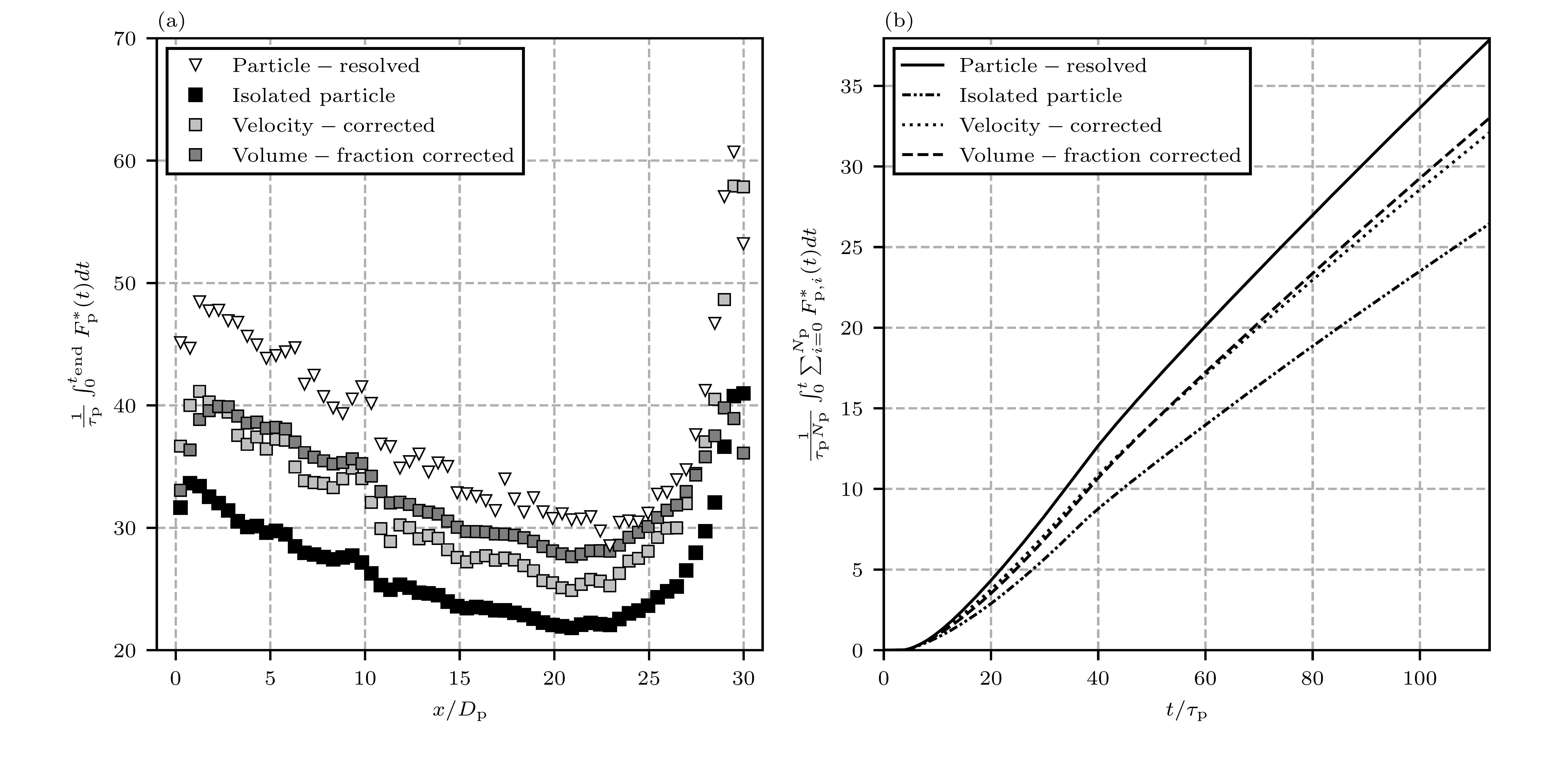}}
	\caption{Spatially averaged particle impulses. (a): Impulses for the particle resolved data and impulses for each particle in the EL simulations, over the full simulation time  $t_\mathrm{end}\approx113\tau_\mathrm{p}$. (b): total impulse as a function of time.}
	\label{fig:EL-impulse}
\end{figure}

\Cref{fig:EL-impulse} shows the impulse over the full simulation time for the particle resolved data and the three EL simulations. The figure also shows how the total impulse (sum over all particles) develops over time. Considering first how the impulse at late time varies with position, we find as expected that the agreement with the particle-resolved data is worst for the isolated particle model, while the velocity-corrected and the volume fraction corrected models are the best in the edge regions and the central regions, respectively. It is worth noting that all three models predict a larger impulse, and thus higher velocity, for the particles near the downstream edge than those further in. It was emphasized in \citet{theofanous2017} that this characteristic is crucial to capture in simulations of shock-accelerated particle layers. The Mach-number dependent quasi-steady drag model ensures that this characteristic is indeed captured by the EL simulations. The impulse is however about 50\% higher for the velocity-corrected model than for the two other models at the very downstream edge. Thus, the dispersive behaviour will be much stronger with this model than the two others, and is likely to be in better agreement with the one that would be observed in both experiments and particle-resolved simulations with moving particles. The source of this improvement is the fluctuation term in the momentum equation. As a test, the velocity-corrected model was run without this term, and the results (not shown), become very similar to the volume-fraction corrected model results. 

For positions near the upstream edge, the impulse prediction is lower than the one obtained with particle-resolved data as input to the drag models. This is related to the strength of the reflected shock wave, which is too weak for all three models.

The development of the impulse over time shows that all models consistently predict an impulse that grows too slowly. Even while the shock-wave is still inside the particle layer ($t<37\taup$), all impulses are too low. A part of the reason for this is likely the negative contribution of the inviscid unsteady force, but it is also partially due to the lower quasi-steady forces. An important consequence of the growing impulse deficit is that if the particles were allowed to move, the particle velocities and consequently their positions will deviate more and more from the correct results with time. In this configuration, the particle velocity deficit at $t=100\taup$ would on average be $30\%$, $15\%$ and $13\%$ for the isolated particle model, the velocity-corrected model and the volume-fraction corrected model, respectively.

\section{Conclusions and future perspectives}
\label{sec:conclusions}
 
This work has compared particle force predictions from standard drag laws to particle forces obtained by means of particle-resolved simulations. Both direct application of the drag laws with volume-averaged particle-resolved simulation fields as input and application of the drag laws within an EL simulation was used to evaluate how well they perform. In addition, two drag law correction models were evaluated with the same approach. 
All three approaches capture the main phenomena occurring in this problem. These are the reflection of the incident shock wave and a gradual increase in the reflected shock wave strength over time, the gradual attenuation of the shock wave as it propagates through the particle layer, and the flow expansion on the downstream edge. 

Without any corrections, the drag laws consistently underpredict forces. With the volume averaged fields as input, the average isolated particle model prediction is just slightly larger than half the average value of the particle-resolved forces. In the EL simulations, the predictions are in slightly better agreement with the particle-resolved forces, but since the improvement is caused by differences in the gas-flow fields, we still conclude that the model is inadequate for use in high-speed dense solid-gas flows. 

The two correction schemes yield improved predictions relative to the isolated particle model predictions. The first correction scheme modifies the velocity that is used as input to the force models, while the second scales three of the force components by a volume-fraction dependent factor. With volume-averaged particle-resolved simulation fields as input, the volume-fraction corrected approach gives the best force predictions. The velocity-corrected approach gives slightly lower forces than the volume-fraction corrected model on average, but is significantly better than the isolated particle model. 

In the EL simulation setting, the situation changes. Here, the difference between the volume-fraction corrected model and the velocity-corrected model forces is smaller, again due to differences in the gas flow fields. The velocity corrected fields are in better agreement with the particle-resolved simulation fields, and thus the input to the force models is better with this approach than with the isolated particle model and the volume-fraction corrected model. The reason for the improvement with the velocity-corrected model is less related to the particle force models than to the introduction of the velocity-fluctuation correlation term in the volume-averaged momentum equation. The velocity-correction model stems from a simple approximation to the flow field within a particle cloud, where separated flow in the particle wakes is assumed to occupy a non-negligible portion of the volume, and thus a velocity-fluctuation correlation arises that must be used together with the velocity correction. The velocity-correlation imposes extra forces on the gas, and these forces depend strongly on the volume fraction gradients. Therefore, the dynamics in the particle-cloud edges are modified, and this improves the EL simulations. 

The velocity fluctuations have clearly been demonstrated to be important in this work as well as in several previous studies, e.g. \cite{regele2014,hosseinzadeh2018,vartdal2018,osnes2019,osnes2019c,shallcross2020}. Here, we have successfully applied a model that assumes instantaneous equilibrium between the fluctuations and the flow around it, as well as fluctuations that are only advected with the particles and not with the gas. In reality, none of these properties are present. Particle-wakes are not generated instantaneously, and once they have been generated, they do not respond immediately to the changes in the surrounding flow. There is a time-delay that should be properly modeled, and an appropriate form of the model could be something resembling the history integrals in the inviscid and viscous unsteady forces. Alternatively, a formulation where mass and momentum exchanges between the separation flow and the surrounding fluid flow are taken into account could be used. Such a model was proposed within a two-fluid formulation by \citet{fox2020}, who also showed that such a model can be hyperbolic. As shown in \labelcref{sec:appA}, the velocity-corrected model behaves poorly under mesh-refinement, a behaviour that is likely due to the instantaneous equilibrium assumption, which influences the model's hyperbolicity. 

In addition to the fluctuations that are "locked" to the particles, which are for example the separated flow behind the particles and the stagnated flow in front of them, the particle wakes generate fluctuations that are advected with the gas. These start out as vortices that form in the shear layers around the particles, and over time become classical turbulent fluctuations. An appropriate model for these fluctuations could resemble $k-\varepsilon$ models, but with fluctuation generation and dissipation terms that are derived especially for high-speed dense gas-solid flows. Such approaches have in fact been attempted for a setting very close to the one studied in this work, first by \citet{vartdal2018} and later by \citet{shallcross2020}. When the fluctuations are advected with the gas flow, the velocity-correlation does not drop as sharply over the downstream edge, and the expansion will not be quite as strong. We believe that the models for the gas-advected and the particle-advected fluctuations should be used simultaneously to achieve a combined model that captures the true behaviour of the fluctuations in the best manner.  

With both model evaluation approaches taken here, a consistent property of the force models is that they predict a force that decays too rapidly following the shock-induced peak force. This can be attributed to the inviscid unsteady force, whose kernel can take negative values between $1.6\leq 2c_\infty t/\Dp \leq 7$, where $c_\infty$ is the ambient speed of sound (see \citet{parmar2009}). The results of this work indicate that the kernels should  either have lower magnitudes or possibly only positive values for the current volume fractions and Mach numbers. Since the inviscid kernels in \citep{parmar2009} were computed by accelerating a sphere in a constant Mach number background flow, their validity in the case of shock-accelerated flow regimes is by no means guaranteed. Indeed, the results of that study indicate that the model accuracy varies significantly with Mach number even for isolated particles. In the incompressible, low-Reynolds number limit, \citet{sangani1991} provided correction factors that scale the inviscid unsteady force for oscillatory flow, but simple scaling is insufficient for capturing the behaviour we have found here. Further studies are needed for establishing inviscid unsteady force kernels for non-negligible Mach numbers and particle volume fractions. Particle-resolved simulations are well suited for this purpose.

The current work has only evaluated the streamwise forces, which, although clearly the most important for the present configuation, are not sufficient for describing the behaviour of shock-accelerated dense solid-gas flows. Lift-forces, or spanwise forces due to local differences in the mean flow direction, are also important in this setting. This is especially true in diverging geometries, such as those found in explosive dispersal problems, where spherical or cylindrical symmetries are immediately broken, something which is most clearly visible by the emergence of particle jets \cite{zhang2001,milne2010}. These asymmetries also introduce volume fraction gradients that are not oriented with the flow direction, and the behaviour of such gradients is not well known. Multiphase flow instabilities in general are an interesting topic in the context of shock-accelerated flow. The relation of multiphase flow instability to physics at the particle scale, as well as the connection between particle force models and instability, are important topics for future studies.

\bibliographystyle{elsarticle-num-names}


\clearpage
\setcounter{figure}{0}
\appendix
\section{Effect of grid-refinement}
\label{sec:appA}
\begin{figure}
	\centerline{
		\includegraphics[]{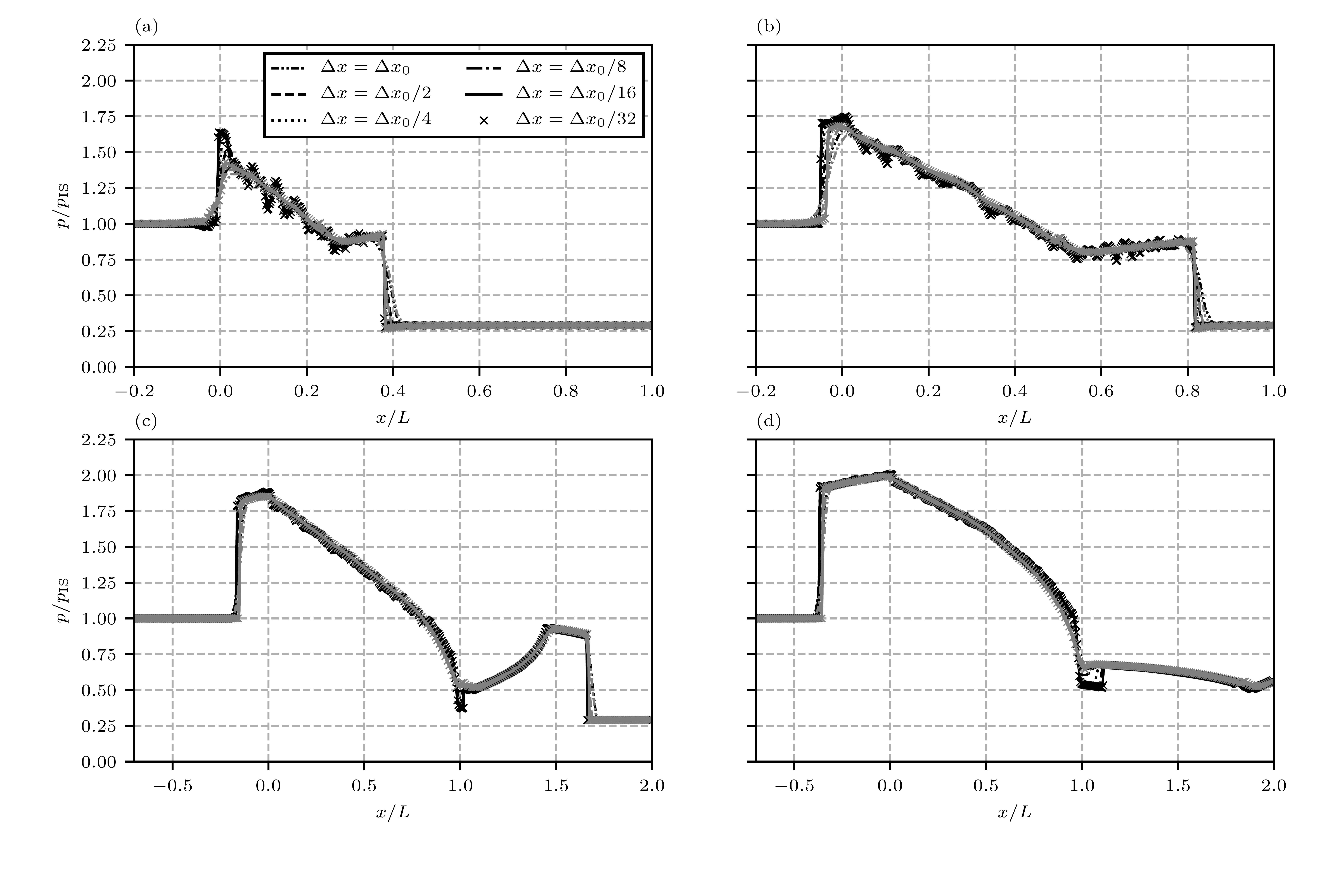}}
	\caption{Mesh refinement test for the isolated particle model, with $\Fiu=\Fvu=0$ at four different times. (a) $t=15\taup$, (b):  $t=30\taup$, (c):  $t=60\taup$, (d):  $t=100\taup$. Black lines and symbols show the results with $\sigma_\mathrm{f}=0.5\Dp$ while grey lines and symbols show results for $\sigma_\mathrm{f}=3\Dp$.}
	\label{fig:meshrefinement_1}
\end{figure}

\begin{figure}
	\centerline{
		\includegraphics[]{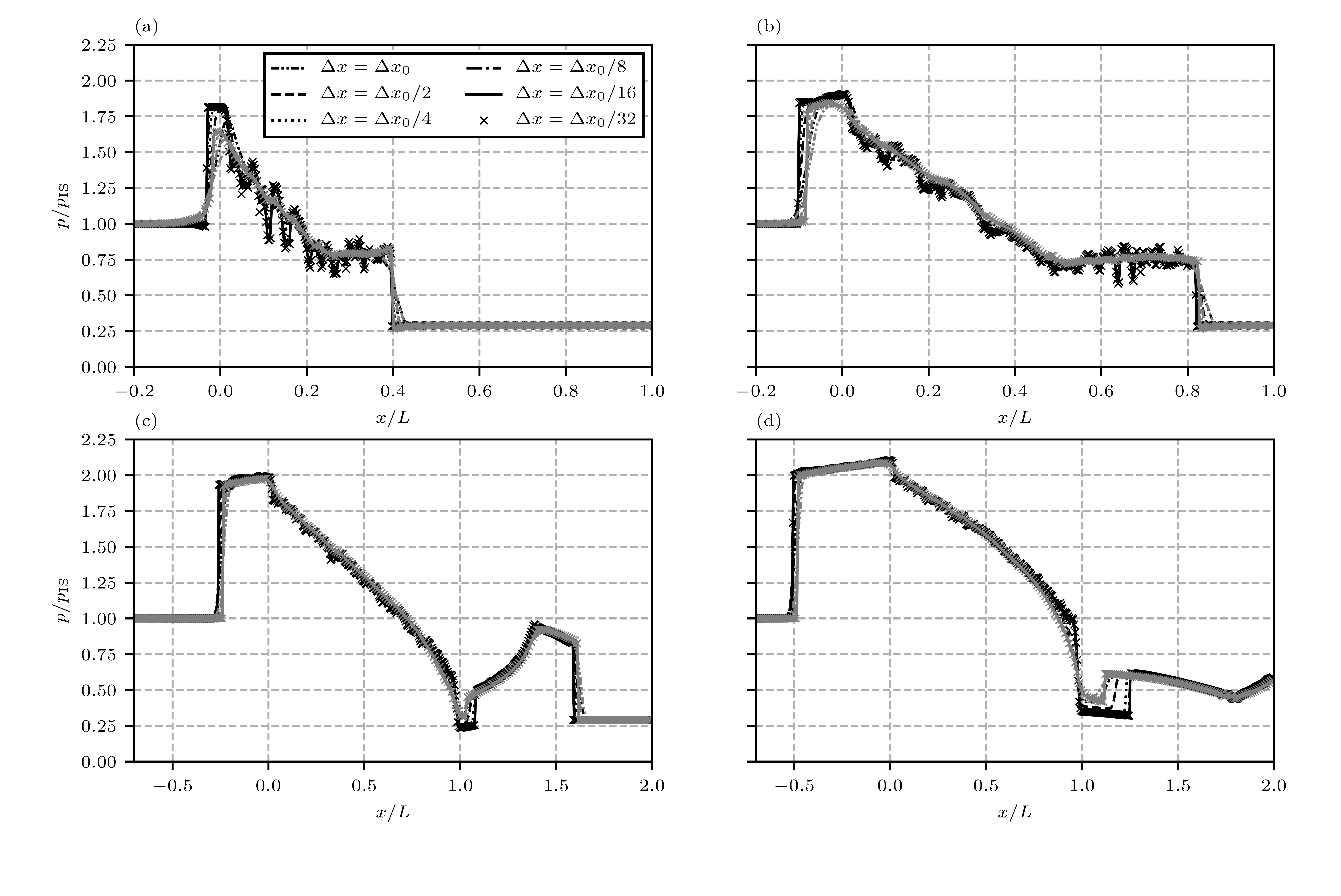}}
	\caption{Mesh refinement test for the velocity-corrected model, with $\Fiu=\Fvu=0$ at four different times. (a) $t=15\taup$, (b):  $t=30\taup$, (c):  $t=60\taup$, (d):  $t=100\taup$. Black lines and symbols show the results with $\sigma_\mathrm{f}=0.5\Dp$ while grey lines and symbols show results for $\sigma_\mathrm{f}=3\Dp$. }
	\label{fig:meshrefinement_2}
\end{figure}

\begin{figure}
	\centerline{
		\includegraphics[]{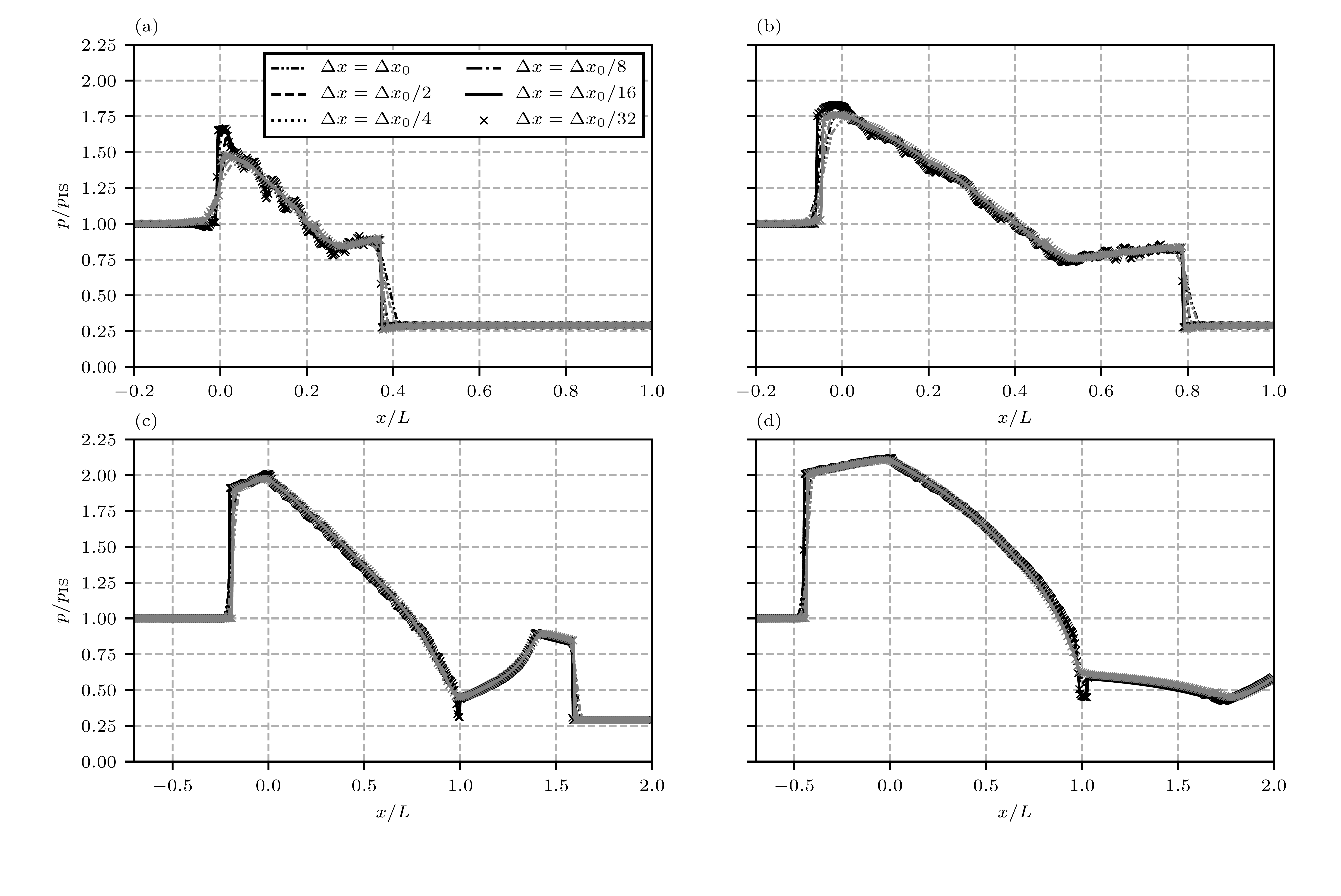}}
	\caption{Mesh refinement test for the volume-fraction corrected model, with $\Fiu=\Fvu=0$ at four different times. (a) $t=15\taup$, (b):  $t=30\taup$, (c):  $t=60\taup$, (d):  $t=100\taup$. Black lines and symbols show the results with $\sigma_\mathrm{f}=0.5\Dp$ while grey lines and symbols show results for $\sigma_\mathrm{f}=3\Dp$. }
	\label{fig:meshrefinement_3}
\end{figure}

\Cref{fig:meshrefinement_1,fig:meshrefinement_2,fig:meshrefinement_3} show the effects of grid-refinement for the three models, with filter-widths $\sigma_\mathrm{f}=0.5\Dp$ and $\sigma_\mathrm{f}=3\Dp$. Here, $\Delta x_0$ corresponds to the width of the volume averaging bins used for the particle-resolved simulations, which is approximately $0.5\Dp$. These results were computed with $\Fiu=\Fvu=0$ since the computation of these terms are the most expensive part of the simulation, but are unlikely to affect the manner in which the solution converges as the grid is refined. The results are reasonably well captured even at the coarsest grids, but there are some minor changes that occur as the grid is refined. The most important of these is that the reflected shock forms more quickly, and that the expansion at the downstream edge is slightly stronger. Additionally, the shock wave(s) sharpen with increasing spatial resolution, as expected.

The figures also show the differences between different filter widths for the volume fraction field for the three models. Especially for the isolated particle model, a smaller filter appears to be advantageous for capturing the expansion flow at the downstream particle cloud edge. For all three models, the expansion becomes stronger with smaller filter widths, which is likely due to a larger magnitude of the volume fraction gradient.

\end{document}